\documentclass{article}
\usepackage{graphicx} 
\usepackage{blindtext}
\usepackage{amsmath}
\usepackage{amsfonts}
\usepackage{amssymb}
\usepackage{amsthm}
\usepackage{graphicx}
\usepackage{tikz-cd}
\usepackage{caption}
\usepackage{float}
\usepackage[colorlinks=true, allcolors=blue]{hyperref}

\newcommand{\Z}{{\mathbb{Z}}}

\newcommand{\beq}{\begin{equation}}
\newcommand{\eeq}{\end{equation}}
\newcommand{\beqn}{\begin{eqnarray}}
\newcommand{\eeqn}{\end{eqnarray}}

\newcommand{\cH}{ {\cal H} }

\newcommand{\RR}{\mathbb{R}}

\newcommand{\lr}[1]{\lfloor #1 \rceil}

\def\[#1\]{%
  \begin{equation}\begin{gathered}#1\end{gathered}\end{equation}%
}
\usepackage{authblk}
\title{Chiral Lattice Gauge Theories \\from Symmetry Disentanglers}
\author[1]{Ryan Thorngren}
\affil[1]{Mani L. Bhaumik Institute for Theoretical Physics, \protect\\ University of California, Los Angeles, CA 90095, USA}
\author[2]{John Preskill}
\affil[2]{Institute for Quantum Information and Matter, \protect\\ California Institute of Technology, CA 91125, USA}
\author[3]{Lukasz Fidkowski}
\affil[3]{Department of Physics, \protect\\ University of Washington, Seattle, WA 98195, USA}
\date{January 2026}

\begin{document}

\maketitle

\begin{abstract}
    We propose a Hamiltonian framework for constructing chiral gauge theories on the lattice based on \textit{symmetry disentanglers}: constant-depth circuits of local unitaries that transform not-on-site symmetries into on-site ones. When chiral symmetry can be realized not-on-site and such a disentangler exists, the symmetry can be implemented in a strictly local Hamiltonian and gauged by standard lattice methods. Using lattice rotor models, we realize this idea in 1+1 and 3+1 spacetime dimensions for $U(1)$ symmetries with mixed 't Hooft anomalies, and show that symmetry disentanglers can be constructed when anomalies cancel. As an example, we present an exactly solvable Hamiltonian lattice model of the (1+1)-dimensional ``3450'' chiral gauge theory, and we argue that a related construction applies to the $U(1)$ hypercharge symmetry of the Standard Model fermions in 3+1 dimensions. Our results open a new route toward fully local, nonperturbative formulations of chiral gauge theories. 
    \end{abstract}
\tableofcontents

\section{Overview}

Chiral gauge theories play a central role in modern physics. A prime example is the Standard Model of particle physics; its fermions transform chirally under the gauge group, and the consistency of the theory relies on highly nontrivial cancellations of ’t Hooft anomalies. But despite decades of effort, a local non-perturbative formulation of chiral gauge theories in 3+1 dimensions remains elusive.

Lattice regularization provides a powerful framework for defining quantum field theories beyond perturbation theory, enabling advances both in numerics \cite{Kronfeld_2012,LGTbeyondSM} and rigorous constructive approaches \cite{gallavotti1985renormalization,bauerschmidt2019introduction}. In favorable cases, one constructs a local lattice Hamiltonian whose long-distance physics flows to a target continuum theory. However, when applied to chiral fermions this program encounters a fundamental obstruction: lattice models with strictly local interactions and strictly on-site symmetries generically produce fermions in vector-like pairs \cite{nielsen1981absence}. This ``fermion doubling'' phenomenon blocks a straightforward lattice realization of chiral gauge theories.

Strategies have been developed to evade fermion doubling, some of which can be understood from the perspective of anomaly inflow.  From this viewpoint, chiral fermions in $D$ spacetime dimensions arise as boundary modes of a gapped system in $D{+}1$ dimensions \cite{Kaplan_1992,kaplan2024chiral}, whose bulk response encodes the boundary anomaly. The bulk system, known as a symmetry-protected topological (SPT) phase in condensed matter language, is gapped and short-range entangled, and, in cases when it is possible to effectively decouple its boundary, the action of the symmetry on this boundary is not-on-site; that is, it cannot be written as a tensor product of local operators.  In the overlap fermion approach \cite{Narayanan_1995,hernandez1999locality}, the massive bulk modes are integrated out and, in some cases where 't Hooft anomalies cancel, the chiral symmetry can be consistently gauged by modifying the fermionic path-integral measure \cite{L_scher_1999}. However, this method does not provide an explicitly local Hamiltonian; we seek a model in a tensor-product Hilbert space with bounded-range interactions and a symmetry that is manifestly on-site. In the symmetric mass generation (SMG) approach \cite{eichten1986chiral} (see also \cite{POPPITZ_2010,Wen_2013,wang2022symmetric,hasenfratz2025symmetric,tong2020notes} for some recent perspectives), one considers a $(D{+}1)$-dimensional slab with finite thickness, where chiral fermions reside on the upper $D$-dimensional boundary, and mirror fermions reside on the lower $D$-dimensional boundary. In some cases where 't Hooft anomalies cancel on each boundary, strong interactions localized on the lower boundary can fully gap the mirror fermions without breaking the chiral symmetry, resulting in a low-energy effective theory with chiral fermions and canceling 't Hooft anomalies. A drawback of SMG is that analyzing it properly requires understanding the strongly-coupled mirror fermion sector. 

Here we propose a different approach, focusing on the lattice realization of chiral symmetry rather than strongly-coupled fermion dynamics. If a lattice symmetry is defined on-site, then it can be consistently gauged by standard lattice methods. Therefore, in a $D$-dimensional theory with chiral symmetry, uncanceled 't Hooft anomalies obstruct realizing the symmetry on-site, but a not-on-site realization of chiral symmetry may be possible nevertheless. If chiral symmetry is realized not-on-site in $D$ dimensions and 't Hooft anomalies cancel, we may ask: Can the not-on-site symmetry be rendered on-site by a local change of basis? If so, the symmetry can be straighforwardly gauged. Hence whether the chiral gauge theory can be constructed hinges on the existence of a \textit{symmetry disentangler}: a constant-depth circuit of local unitaries that transforms the not-on-site symmetry action into an on-site one.

Recent work shows there are obstructions to gauging symmetries on the lattice that have no analogue in the continuum~\cite{shirley2025anomalyfreesymmetriesobstructionsgauging,tu2025anomaliesglobalsymmetrieslattice,czajka2025anomalieslatticehomotopyquantum}. Therefore, anomaly cancellation in quantum field theory does not, in general, guarantee the existence of a corresponding lattice construction or of a symmetry disentangler. Nevertheless, we will show that symmetry disentanglers can be explicitly constructed for several physically interesting examples. In these cases, lattice models with chiral symmetry can be realized.

As a first example, consider $D=2$.  It is already known how to construct lattice models describing a (1+1)D compact boson with on-site $U(1)_V$ and not-on-site $U(1)_A$ symmetry sharing a mixed 't Hooft anomaly, the famous chiral anomaly \cite{Cheng_2023}. Building on this observation, we consider a stack of $N$ such bosonic systems, with the full symmetry group $G^N$, where $G= U(1)_A \times U(1)_V$. Given a $U(1)$ subgroup of $G^N$ with canceling 't Hooft anomalies, we show that a symmetry disentangler can be explicitly constructed. This construction extends recent work by Seifnashri and Shirley on disentangling anomaly-free \emph{discrete} symmetries in (1+1)D \cite{shirley2025,Seifnashri_2024}.

By this scheme we obtain a local, exactly solvable (1+1)D lattice model realizing the so-called ``3450 theory,'' consisting of two left-moving Weyl fermions of charge 3 and 4 and two right-moving Weyl fermions of charge 5 and 0. The resulting model has a tensor-product Hilbert space, albeit with infinite-dimensional rotor degrees of freedom, and a $U(1)$ symmetry that can be gauged to yield a chiral gauge theory. To our knowledge, this is the first solvable Hamiltonian model for the 3450 theory. Previous constructions include solvable Euclidean lattice models~\cite{berkowitz2024exact,morikawa2024latticeformulation2du1}, or Hamiltonian approaches based on symmetric mass generation that are validated numerically~\cite{Wang_2019,Zeng_2022}.

\begin{figure}
    \centering
    \includegraphics[width=10cm]{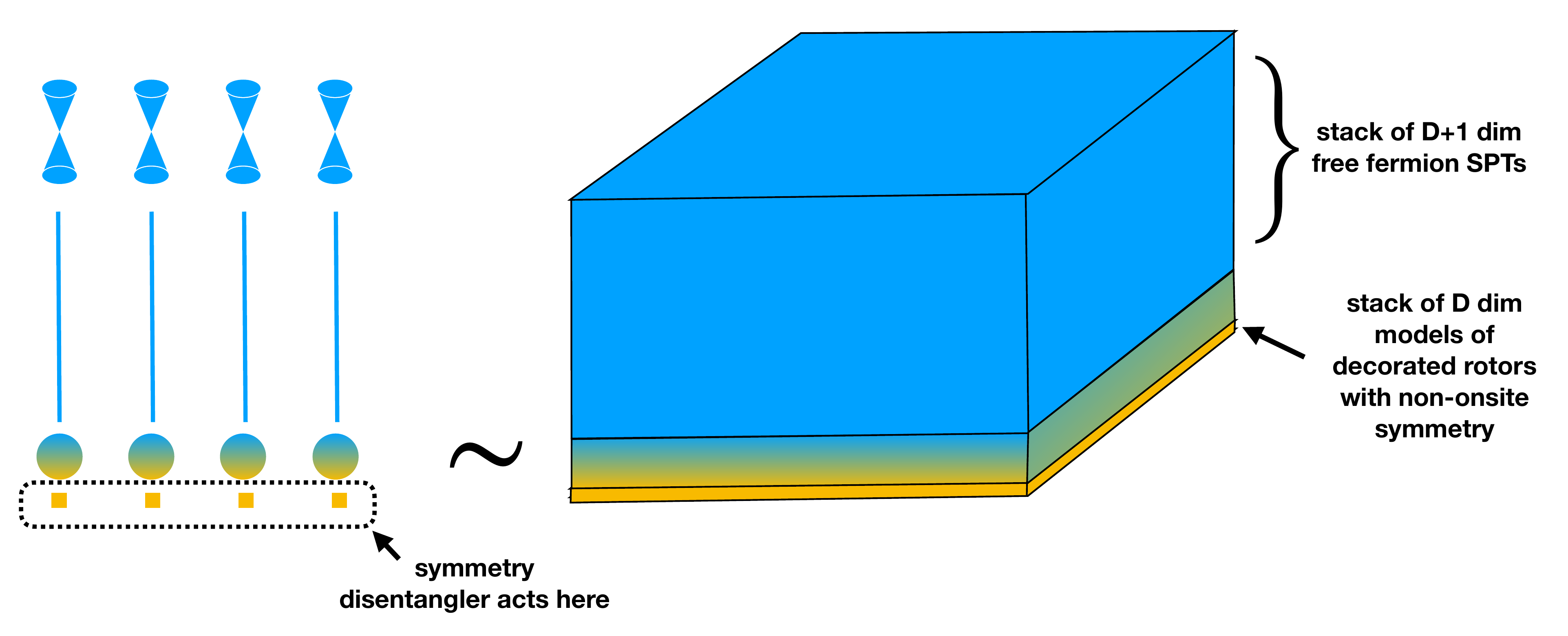}
    \caption{Using a symmetry disentangler to construct a chiral theory.  Within a stack of free-fermion $G$-SPTs (blue), the lower boundary of each such SPT is $G$-symmetrically gapped (blue-yellow region) against a corresponding $D$ dimensional model (yellow) with the same anomaly but with a not-on-site realization of $G$, resulting in a topologically trivial lower boundary.  Then the only remaining low-energy degrees of freedom are those on the upper boundaries of the free-fermion SPTs (blue), yielding a chiral model. The final step is to make an anomaly free subgroup $G' \subset G \times \ldots \times G$ on-site by applying a symmetry disentangler that acts across the stack of $D$-dimensional Hilbert spaces. }
    \label{fig:3450-figureA}
\end{figure}

In $D=4$ dimensions, we can pursue an analogous strategy. We write down anomalous $G=U(1)_V \times U(1)_A$ symmetries and show that if we take a stack of such systems with canceling 't Hooft anomalies for some subgroup $G' \cong U(1) \subset G\times \ldots \times G$ of the full symmetry group of the stack, then a $G'$ symmetry disentangler again exists.  Conjugating the Hamiltonian by such a disentangler renders $G'$ on-site without affecting the Hamiltonian's low-energy spectrum, since constant-depth circuits preserve spectral properties. Once the $G'$ symmetry is realized on-site, it can be straightforwardly gauged.

This strategy is complicated by our inability to find exactly solvable Hamiltonians that commute with the not-on-site symmetry $G$ of the lattice Hilbert space.  To make progress, we combine our construction with the domain wall fermion framework, as illustrated in Fig.~\ref{fig:3450-figureA}.  Concretely, each not-on-site $D$-dimensional model in our stack is coupled to the lower boundary of a $D{+}1$-dimensional slab hosting a free-fermion $G$-SPT \cite{Qi_2008,Kitaev_2009,Ryu_2010}.  The free-fermion SPT is chosen to have a boundary anomaly matching that of the not-on-site $D$-dimensional model. This anomaly matching plausibly allows the not-on-site model and the lower boundary of the SPT to be gapped out together while preserving the $G$ symmetry. Then the only remaining low-energy degrees of freedom are those on the upper boundaries of the free fermion SPTs, yielding a model of anomalous fermions with not-on-site $G$ symmetry.

The $(D{+}1)$-dimensional SPT has on-site $G$ symmetry; hence the not-on-site action of the symmetry on the slab arises entirely from the $D$-dimensional models we have coupled to the lower boundary. If we consider a stack of these slabs with symmetry $G \times \cdots \times G$ and identify an anomaly-free subgroup $G' \cong U(1)$, then we can apply our symmetry disentangler, supported near the lower boundary. The disentangler renders the action of $G'$ on-site on the full microscopic Hilbert space, while leaving the gapless chiral modes living on the upper boundary untouched. Once the symmetry is realized on-site at the microscopic level, it can be gauged by standard lattice methods, yielding a chiral gauge theory with canceling 't Hooft anomalies.

This picture resembles the SMG scenario, but with an important difference: we don't need to invoke strong interactions to gap out the lower boundary, assuming that we can construct the symmetry disentangler. Gapping out each free-fermion SPT boundary together with its corresponding not-on-site model \textit{does} requires interactions, but these act only within each layer separately, do not couple across the different SPTs in the stack, and have nothing to do with anomaly cancellation among these SPTs. In the models where we construct symmetry disentanglers, both $G$ and the free fermion SPT boundaries are simple enough that we can provide a heuristic to perform the gapping. For SMG, in contrast, strong-coupling across the SPTs is essential for gapping out the lower boundary.  In section \ref{sec:in-flow} below we describe a variant of our construction that uses the anomaly in-flow point of view and even more closely resembles an exactly solvable version of the SMG scenario.

One might wonder why the SPTs are necessary for our construction. The $D$-dimensional systems we consider already have $G'$ symmetry realized not-on-site, and for the specific examples we study in which 't Hooft anomalies cancel, a symmetry disentangler can be applied directly to these systems (together with suitable ancilla degrees of freedom) without introducing any SPTs at all.

The role of the ($D$+1)-dimensional SPTs is instead to provide a controlled mechanism for realizing the desired chiral infrared theory from a microscopic Hamiltonian. Introducing the SPTs converts the problem of ``emerging'' the chiral theory into the problem of gapping the original $D$-dimensional system against the lower boundaries of the SPT slabs, while preserving the $G'$ symmetry and leaving the upper boundary gapless. By providing a concrete mechanism to address this gapping problem, we arrive at a well-defined lattice realization of the chiral theory.

We argue below that such constructions can reproduce the $U(1)_Y$ hypercharge assignments of a single fermion generation in the Standard Model. In this case the quark sector by itself and the lepton sector by itself both have uncanceled $U(1)_Y^3$ 't Hooft anomalies, but when the two sectors are combined the anomalies cancel and a symmetry disentangler exists that maps the not-on-site $U(1)_Y$ to an on-site symmetry that can be gauged. While a complete Hamiltonian realization of the full Standard Model has not yet been achieved, our results suggest a promising route: the symmetry disentangler enables us to achieve chiral fermion content under $U(1)_Y$ without resorting to uncontrolled strong dynamics in the mirror sector.

More broadly, our work sharpens a Hamiltonian viewpoint on anomalies and gaugeability: in the class of lattice symmetries we study, anomaly cancellation aligns with the existence of a constant-depth circuit that renders the symmetry on-site, and hence gaugeable. We expect this ``disentangling'' perspective to be useful beyond the specific models considered here, by connecting anomaly inflow, lattice chiral symmetries, and the construction of symmetric gapped boundaries within a single analytically tractable framework.

\subsection{Towards the Standard Model}

Here we make more comments on the applicability of our method to the hypercharge sector of the Standard Model. We find that the hypercharge assignments for one generation of quarks and leptons are fortuitously chosen to allow the construction of the symmetry disentangler. For this construction to work, a sterile neutrino, which carries no Standard Model charge, must be added to the fermion content. It is an intriguing question whether this feature of the construction is a mere technicality or has a deeper significance.

The Standard Model is a gauge theory with gauge group $G = SU(3) \times SU(2) \times U(1)_Y$ containing fermion matter with chiral charges under $SU(2)$ and hypercharge $U(1)_Y$. We can express this content in terms of three generations of quarks and leptons with each generation consisting of a left-handed Weyl fermion transforming under the representation
\[SU(3) \times SU(2) \times U(1)_Y: (3,2)^1 +(\bar 3,1)^{-4} + (\bar 3,1)^2 + (1,2)^{-3} + (1,1)^6\]
where we have scaled the hypercharges (written in the superscripts) by 6 so the smallest hypercharge is 1. Focusing on just the hypercharges, the fermion content separates into three groups of quarks (one for each color) and one group of leptons, with $U(1)_Y$ hypercharges
\[\label{eqnSMhypercharges}\begin{matrix}
    \text{quark hypercharges ($\times 3$)}  & -4 & 2& 1 & 1 \\
    \text{lepton hypercharges} & 6 & 0& -3 & -3 
\end{matrix}\]
Here the first two columns are $SU(2)$ singlets, and the last two columns form $SU(2)$ doublets. Note that we have included a a sterile neutrino (a Weyl fermion with zero charge). There are 16 Weyl fermions with these $U(1)_Y$ charge assignments in each generation of quarks and leptons.

We observe that both the quark hypercharges and the lepton hypercharges are linear combinations of the charge vectors
\[\label{eqndiraccharges}\begin{matrix}
    U(1)_V & +1 & -1 & 0 & 0 \\
    U(1)_A & +1 & +1 & -1 & -1
\end{matrix}\]
Four left-handed Weyl fermions with this $U(1)_V \times U(1)_A$ symmetry have an anomaly.\footnote{Note a $\pi$ rotation in $U(1)_A$ corresponds to fermion parity, so this symmetry combines with the Lorentz symmetry as $(\text{Spin} \times U(1)_A)/\Z_2 \times U(1)_V$. We give a careful account of the anomaly in Appendix \ref{appchiralanom}.} It turns out that this is precisely the minimal such anomaly we can realize on the lattice with not-on-site symmetries via our approach.

In particular, we are able to define lattice symmetries $U(1)_{V_i}$, $U(1)_{A_i}$, $i = 1,2,3,4$, and a subgroup $U(1)_i \subset U(1)_{V_i} \times U(1)_{A_i}$ with the charge assignments of each of the groups \eqref{eqnSMhypercharges} above, as well as a symmetry disentangler for the diagonal $U(1)_Y \subset \prod_{i=1}^4 U(1)_i$. Thus we can apply the general scheme outlined above.

That the charges must be groupable in this way for our construction to work is a consequence of how we write down not-on-site symmetries on the lattice. We are not able, for instance, to express a $U(1)$ symmetry which would have the same anomaly as a single charge-1 Weyl fermion, since such anomaly has a gauge-gravitational contribution $A\ {\rm Tr} R\wedge R$, which seems impossible to realize in a local lattice model \cite{Fidkowski_Xu}. We stress that there are even cases where all 't Hooft anomalies vanish, but the charges cannot be grouped as in our construction.  For example, five left-handed Weyl fermions with charges 1,5,9,$-7$,$-8$ also have vanishing $U(1)_Y^3$ and vanishing mixed gauge-gravitational anomalies, but cannot be expressed in groups as above because there are an odd number of them.

The paper is organized as follows. In Section \ref{sec:1+1d} we describe the construction of chiral $U(1)$ gauge theories in 1+1D. We show how to convert an exactly-solvable Villain model of a compact boson to one in a local rotor Hilbert space with on-site $U(1)_V$ and not-on-site $U(1)_A$ symmetry sharing the usual chiral anomaly. We then describe how to construct symmetry disentanglers for anomaly-free combinations of these models, allowing us to produce arbitrary chiral $U(1)$ gauge theories of compact bosons, subject only to anomaly-vanishing. These models may be converted to models of Dirac fermions by familiar Jordan-Wigner transformations.

In Section \ref{sec:3+1d} we describe the 3+1D constructions. We review the construction of bosonic $U(1)_V \times U(1)_A$ symmetry with a certain mixed anomaly. We describe how to disentangle anomaly-free combinations of these symmetries, which can then be gauged. The generalization to fermions is not so straightforward, but we are able to do it, finding symmetries whose anomaly corresponds to the four Weyl fermions in eq.~\ref{eqndiraccharges}. We also show how to disentangle anomaly-free combinations of the fermionic $U(1)$ symmetries.

In Section \ref{sec:in-flow} we connect our native $D$-dimensional constructions to an in-flow picture in $D+1$-dimensions, and sketch a construction of a model giving rise to the $U(1)$ hypercharge sector of the Standard Model.

\textit{Acknowledgements:} We thank Qing-Rui Wang for discussions of his work \cite{Wang_2018}, which we make use of below. RT is grateful to Lei Gioia for related collaboration. LF acknowledges Cenke Xu and Carolyn Zhang for related collaborations. We also thank Theodore Jacobson for discussions. RT is supported by the Bhaumik Presidential Term Chair at the Mani L. Bhaumik Institute for Theoretical Physics.  JP acknowledges support from the U.S. Department of Energy QuantISED
program through the theory consortium “Intersections of QIS and Theoretical Particle Physics” at Fermilab,  the U.S. Department of Energy Office of High Energy Physics (DE-SC0018407), and the Institute for Quantum Information and Matter, an NSF Physics Frontiers Center (PHY-2317110). LF is supported by NSF DMR- 2300172.

\section{1+1 dimensions}\label{sec:1+1d}

\subsection{A lattice rotor model with not-on-site $U(1)_V$ and $U(1)_A$ symmetries}\label{subsec1drotormodel}

Working in the Hilbert space of a lattice model of rotors in $1$ spatial dimension, we construct not-on-site actions of $U(1)_V$ and $U(1)_A$ symmetries, which are characterized by a mixed anomaly.  Ultimately our Hilbert space will be a tensor product of rotors, but we begin with a Villain formulation.  We imagine our system as living on a large but finite ring.  Thus, let $\phi_j$ be a real-valued variable at site $j$, and let $n_{j-1,j}$ be an integer-valued variable.  The Villain Hilbert space consists of wavefunctions $\Psi(\{\phi,n\})$ invariant under the Villain transformations
\begin{align} \label{Villain_condition}
\phi_j &\rightarrow \phi_j + k_j \nonumber\\
n_{j-1,j} &\rightarrow n_{j-1,j} + k_j-k_{j-1}
\end{align}
for integer $k_j$.  One can think of this Hilbert space as the result of gauging the $\Z \subset \RR$ shift symmetry $\phi_j \mapsto \phi_j +k$.  In this interpretation $n$ is the gauge field, and the Villain condition is Gauss's law.

We then define
\begin{align}\label{eqnvectorvillain}
Q^{\text{Vil}}_V = \frac{i}{2\pi} \sum_j \frac{d}{d\phi_j},
\end{align}
and
\begin{align}\label{eqnaxialvillain}
Q^{\text{Vil}}_A &= \sum_j(\phi_j - \phi_{j-1}- n_{j-1,j}).
\end{align}
Since $\exp(2\pi i Q^{\text{Vil}}_V) = 1$ and $\exp(2\pi i Q^{\text{Vil}}_A) = 1$, these generate two $U(1)$ symmetries, $U(1)_V$ and $U(1)_A$.  These have a mixed 't Hooft anomaly, which can be seen by performing a $2\pi$ $U(1)_A$ rotation in a finite interval $I$. In particular, acting with the operator
\begin{align}
\exp\left(2\pi i \sum_{j = 1}^n(\phi_j - \phi_{j-1}- n_{j-1,j})\right) = \exp(-2\pi i\phi_0)\exp(2\pi i\phi_n)
\end{align}
we find endpoint operators $e^{2\pi i \phi_j}$ carry $U(1)_V$ charge 1. We think of this as a charge pump of a unit $U(1)_V$ charge from one endpoint to the other, and is characteristic of the unit mixed $U(1)_V \times U(1)_A$ anomaly for 1+1D bosons.\footnote{This endpoint charge does not depend on the choice of truncation. Any other truncation of the generator $Q_A^\text{Vil}$ differs from this one by two endpoint operators $O_1$, $O_2$ which must be local, gauge invariant, and such that $O_1 + O_2$ commutes with $U(1)_V$. Since they are local, $O_1$ and $O_2$ individually commute with $U(1)_V$, so the new endpoint operators such as $e^{- 2\pi i (\phi_0 + O_1)}$ have the same charge.} It is a lattice analogue of the familiar spectral flow which produces a $U(1)_V$ charge upon increasing the $U(1)_A$ flux by $2\pi$ \cite{manton1985schwinger,CallanHarvey_vortex}.

The following purely quadratic Hamiltonian in the $\phi_j$:
\begin{align}\label{eq:Luttinger}
H^{\text{Vil}}_{\text{Luttinger}}=\sum_j \left(-\frac{U_0}{2} \frac{d^2}{d\phi_j^2} + \frac{J_0}{2}(\phi_j - \phi_{j-1} - n_{j-1,j})^2\right)
\end{align}
commutes with both $Q_V^{\text{Vil}}$ and $Q_A^{\text{Vil}}$, and is exactly solvable, since it is just a Gaussian problem in the bosons.  It describes a $c=1$ compact boson CFT/Luttinger liquid, as shown in section $4.1$ of ref \cite{Cheng_2023}, with the marginal Luttinger parameter determined by $J_0/U_0$.  The $U(1)_V$ and $U(1)_A$ symmetries act as the momentum and winding symmetries respectively.  The essential idea behind the exact solution of $H_{\text{Luttinger}}$ is to use the Villain condition to push all the non-zero $n$ to a single link and interpret the result as a boundary condition for $\phi$.  The result is coupled harmonic oscillators with a twisted periodic boundary condition.

To get operators on a tensor product Hilbert space, as opposed to the Villain Hilbert space, we define the Villain disentangler
\begin{align}\label{eqnvildisentangler}
C|\{\phi_j,n_{j-1,j}\}\rangle = |\{\phi_j, n_{j-1,j}+\lr{\phi_{j-1} - \phi_j}\}\rangle.
\end{align}
Here $\lr{x}$ is the integer closest to the real number $x$.  In terms of the usual floor function, $\lr{x} = \lfloor x+\frac{1}{2} \rfloor$.  $C$ is a locality-preserving unitary from the Villain Hilbert space to a tensor product one of decoupled periodic $\phi_j$ rotors (invariant under simple integer shifts of $\phi_j$) and $n_{j-1,j}$ variables.  The fact that $C$ is locality preserving can be seen from the expression $C = \prod_j C_j$, where
\begin{align}\label{eqnvillaindisentangler}
C_j = e^{i \lr{\phi_{j-1}-\phi_j} \,\chi_{j-1,j}}
\end{align}
where $\chi_{j-1,j} \in [0,2\pi)$ is the variable dual to $n_{j-1,j}$, in the sense that $\exp(i\chi)|n\rangle = |n+1\rangle$.\footnote{One could attempt to view $C$ as a unitary acting on a big tensor product Hilbert space of states $|\{\phi_j,n_{j-1,j}\}\rangle$, not necessarily those satisfying the Villain condition, in which case eq.~\ref{eqnvillaindisentangler} would be a constant-depth circuit of local unitaries.  However, the states satisfying the Villain condition are not normalizable inside such a large Hilbert space.  Nevertheless, the expression in eq.~\ref{eqnvillaindisentangler} is sufficient for showing that $C$ preserves locality.} Despite the discontinuity in the $\lr{x}$ function, $C$ is a well-defined bounded operator between the two Hilbert spaces.  The fact that our Hilbert space is built from rotors rather than qudits means that natural Hamiltonians on our Hilbert spaces are unbounded operators.  The discontinuity in $\lr{x}$ then introduces a subtlety related to the domains of these un-bounded operators, discussed more in section \ref{sec:discussion}.

Conjugating $Q_V^{\text{Vil}}$ and $Q_A^{\text{Vil}}$ by $C$ we then obtain $Q_V = CQ_V^{\text{Vil}}C^{-1}$ and $Q_A = CQ_A^{\text{Vil}}C^{-1}$ on a tensor product Hilbert space.  Explicitly,
\begin{align}
Q_V = \frac{i}{2\pi} \sum_j \frac{d}{d\phi_j},
\end{align}
and
\begin{align}
Q_A|\{\phi_j,n_{j-1,j}\}\rangle = \sum_j\left(\phi_{j}-\phi_{j-1} -\lr{\phi_{j}-\phi_{j-1}} - n_{j-1,j} \right)|\{\phi_j,n_{j-1,j}\}\rangle.
\end{align}
The Hamiltonian 
\begin{align}\label{eqntensorluttingerham}
H_{\text{Luttinger}}=C H^{\text{Vil}}_{\text{Luttinger}} C^{-1}
\end{align}
is invariant under $Q_V$ and $Q_A$ and realizes a Luttinger liquid in a tensor product Hilbert space with $Q_V$ and $Q_A$ acting as compact momentum and winding respectively.  In Appendix \ref{appdisentanglingvillaingaugetheory} we briefly investigate a related disentangler for Villain $U(1)$ gauge theory.

We can make these formulae more compact with cochain notation (see Appendix \ref{appcochains} for a crash course).  We have that $\phi \in C^0(S^1, \RR)$ and $n \in C^1(S^1, \Z)$, so that the action of $Q_A$ can be compactly expressed as
\begin{align}
Q_A |\phi,n\rangle &= \int \left(d\phi-\lr{d\phi} - n\right)\,|\phi,n\rangle.
\end{align}
Since the field $n$ is not involved in $Q_V$ we can remove it while maintaining the mixed anomaly.  That is, we could define the following simpler operators $\bar{Q}_V$ and $\bar{Q}_A$ on just the Hilbert space of $\phi$ rotors:
\begin{align}
\exp(-2\pi i \beta \bar{Q}_V)|\phi\rangle &=|\phi+\beta\rangle\\
\bar{Q}_A|\phi\rangle &= \int(d\phi-\lr{d\phi})|\phi\rangle.
\end{align}
$\bar{Q}_V$ and $\bar{Q}_A$ still have the mixed anomaly described above.  This is the picture that we generalize to $3+1$ dimensions.

\subsection{Anomaly-cancellation implies on-site-ability}

Let us now take $L$ stacked copies of the above rotor Hilbert space, with rotors $\phi^{\alpha}$, $\alpha=1,\ldots, L$. Given integers $q_V^{\alpha}, q_A^{\alpha}$, we construct
\begin{align}\label{eq:Qmatterdef}
Q_{\text{matter}} = \sum_{\alpha=1}^L \left(q_V^{\alpha} {Q}_V^{\alpha} +q_A^{\alpha} {Q}_A^{\alpha}\right)
\end{align}
as well as ${\bar Q}_\text{matter}$ by replacing $Q_V$ and $Q_A$ by their barred versions.  The sum of Luttinger Hamiltonians defined above, with Luttinger parameters all chosen to be $K=1$, gives a free boson field theory that can be fermionized, at the field theory level, to give $L$ right movers and $L$ left movers.  Assuming $q_V^\alpha = q_A^\alpha \text{ mod } 2$, $Q_{\text{matter}}$ generates a $U(1)$ symmetry in this fermionic theory, with left and right movers carrying charges $\frac{1}{2}(q^{(\alpha)}_V \pm q^{(\alpha)}_A)$ respectively.  For appropriate choices of $q_V^{\alpha}$ and $q_A^{\alpha}$, $Q_{\text{matter}}$ will be a non-anomalous $U(1)$ symmetry.  Specifically, the anomaly-free condition is that
\begin{align}
\sum_{\alpha=1}^L\left((q^{\alpha}_V + q^{\alpha}_A)^2-(q^{\alpha}_V - q^{\alpha}_A)^2\right) = 0
\end{align}
or equivalently
\begin{align} \label{eq:anomaly_free}
\sum_{\alpha=1}^L q^{\alpha}_V q^{\alpha}_A = 0.
\end{align}
One such anomaly-free choice is $L=2$, $q_V^{1} = 3, q_A^{1} = 3$, $q_V^{2}=9, q_A^{2}=-1$, which corresponds to the $3450$ theory.

Assuming the anomaly-free condition (eq.~\ref{eq:anomaly_free}) we will now construct a constant-depth circuit of local unitaries such that conjugating by it makes the non-anomalous $U(1)$ symmetry on-site.  In fact, it is enough to do this for ${\bar Q}_{\text{matter}}$, since the $Q_V^\alpha$ and $Q_A^\alpha$ all already act on-site on $n'$.  Our construction is inspired by that of ref. \cite{shirley2025}, adapted to the continuous symmetry group $U(1)$.  First, we introduce new ancilla rotors $\theta_j \in \RR$ ($\theta \in C^0(S^1, \RR)$), with all wavefunctions invariant under integer shifts of $\theta_j$.  We define
\begin{align} \label{eq:defQancilla}
Q &= {\bar Q}_{\text{matter}} + Q_{\text{ancilla}} \\
Q_{\text{ancilla}} &= \frac{i}{2\pi} \sum_j\frac{d}{d\theta_j},
\end{align}
i.e. the ancillas are charged under $Q$.  Our disentangler $W$ will then make $Q$ on-site.  It is defined by
\begin{align}
W|\phi^\alpha, \theta\rangle = W_{\phi^{\alpha}, \theta }\,|\phi^{\alpha} + q^{\alpha}_V \theta, \theta\rangle
\end{align}
where the phase $W_{\phi^{\alpha}, \theta}$ is defined by
\begin{align} \label{eq:defW}
W_{\phi^{\alpha}, \theta} = \exp\left[-i \int \theta\left( \sum_{\alpha}q_A^{\alpha} \lr{d\phi^{\alpha} +q_V^{\alpha}d\theta}  - d\lr{\sum_{\alpha} q_A^{\alpha} \phi^{\alpha}} \right) \right].
\end{align}
The above integral is really a sum over $1$-simplices, i.e. edges, in $S^1$, and the above formula can be re-written as a product of local unitaries over edges.  An important point is that these unitaries are well defined in the rotor Hilbert space, which follows if we can show that the integrand in the expression above shifts by an integer under individual integral shifts of $\phi^\alpha$ and $\theta$.  To see this, let us first consider shifts $\phi^\alpha \rightarrow \phi^\alpha + m^\alpha$, where $m^\alpha \in C^0(S^1,\Z)$.  Since $\lr{x+m} = \lr{x}+m$ for integral $m$ and any $x$, the factor of $m^\alpha$ cancels between the two terms in the integrand.  Now consider shifts $\theta \rightarrow \theta + m$.  By the anomaly free condition,
\begin{align}
\sum_\alpha q_A^\alpha\lr{d\phi^\alpha+q_V^\alpha(d\theta+dm)} &= \sum_\alpha q_A^\alpha\lr{d\phi^\alpha+q_V^\alpha d\theta} + \sum_\alpha q_A^\alpha q_V^\alpha \,dm \\ &= \sum_\alpha q_A^\alpha\lr{d\phi^\alpha+q_V^\alpha d\theta}
\end{align}
so that the variation of the integrand in eq.~\ref{eq:defW} is 
\begin{align}
-i m \left( \sum_{\alpha}q_A^{\alpha} \lr{d\phi^{\alpha} +q_V^{\alpha}d\theta}  - d\lr{\sum_{\alpha} q_A^{\alpha} \phi^{\alpha}} \right) \in \Z
\end{align}
as desired.  Hence $W$ can be expressed as a product of commuting local unitaries, which can trivially be made into a depth $2$ circuit by staggering even and odd edges.

Now let us see that $W$ actually disentangles the symmetry.  Define $U(\beta) \equiv \exp(i\beta Q)$.  We have
\begin{align}
U(\beta) \equiv \exp(i\beta Q)=\exp(i\beta {\bar{Q}}_{\text{matter}}) \otimes \exp(i\beta Q_{\text{ancilla}}).
\end{align}
We now check that
\begin{align}
    W^\dagger\left(\exp(i\beta {\bar{Q}}_{\text{matter}}) \otimes \exp(i\beta Q_{\text{ancilla}})\right)W={\bf{1}} \otimes \exp(i\beta Q_{\text{ancilla}})
\end{align}
which implies that we have made $Q$ on-site.  We will check the equivalent condition
\begin{align}
    W\left({\bf{1}} \otimes \exp(i\beta Q_{\text{ancilla}})\right)W^\dagger=\exp(i\beta {\bar{Q}}_{\text{matter}}) \otimes \exp(i\beta Q_{\text{ancilla}}).
\end{align}
We have:
\begin{align}
W&\left({\bf{1}} \otimes \exp(i\beta Q_{\text{ancilla}})\right)W^\dagger|\phi^{\alpha},\theta\rangle = \\
W&\left({\bf{1}} \otimes \exp(i\beta Q_{\text{ancilla}})\right)\cdot A\cdot|\phi^{\alpha}- q_V^{\alpha}\theta,\theta\rangle= \\
W&\cdot A \cdot |\phi-q_V^{\alpha}\theta, \theta+\beta\rangle = \\
A' &\cdot A \cdot |\phi+q_V^{\alpha}\beta, \theta+\beta\rangle
\end{align}
where the phase factors $A$ and $A'$ are:
\begin{align}
A=\exp\left[i \int \theta \left(\sum_\alpha q_A^{\alpha} \lr{d\phi^{\alpha}} - d\lr{\sum_{\alpha} q_A^{\alpha} \phi^{\alpha}}  \right) \right]
\end{align}
and
\begin{align}
A'=\exp\left[-i \int (\theta+\beta) \left(\sum_\alpha q_A^{\alpha} \lr{d\phi^{\alpha}} - d\lr{\sum_{\alpha} q_A^{\alpha} \phi^{\alpha}}  \right) \right].
\end{align}
We have again used the anomaly-free condition $\sum_{\alpha} q_V^{\alpha} q_A^{\alpha}=0$ to simplify the expression for $A'$.  Thus we finally obtain:
\begin{align}
W&\left({\bf{1}} \otimes \exp(i\beta Q_{\text{ancilla}})\right)W^\dagger|\phi^{\alpha},\theta\rangle = \\
&\exp\left(-i\,\beta\int\sum_{\alpha} q_A^{\alpha} \lr{d\phi^{\alpha}} \right)|\phi+q^{\alpha}\beta, \theta+\beta\rangle.
\end{align}
This is precisely the exponentiated action of our symmetry $Q$: the $\theta$ get rotated by angle $\beta$ (from the $Q_{\text{ancilla}}$ part of $Q$), the $\phi^{\alpha}$ get rotated by angles equal to $q_V^{\alpha} \beta$ (from the $\sum_{\alpha} q_V^{\alpha} {\bar{Q}}_V^{\alpha}$ part of ${\bar{Q}}_{\text{matter}}$, eq.~\ref{eq:Qmatterdef}), and the phase $\exp\left(-i\,\beta\int\sum_{\alpha} q_A^{\alpha} \lr{d\phi^{\alpha}} \right)$ (from the $\sum_{\alpha} q_A^{\alpha} {\bar{Q}}_A^{\alpha}$ part of ${\bar{Q}}_{\text{matter}}$, eq.~\ref{eq:Qmatterdef}) is imprinted on the configuration.  We thus conclude that
\begin{align}
    W\left({\bf{1}} \otimes \exp(i\beta Q_{\text{ancilla}})\right)W^\dagger=\exp(i\beta {\bar{Q}}_{\text{matter}}) \otimes \exp(i\beta Q_{\text{ancilla}}),
\end{align}
i.e.
\begin{align}
    W^\dagger \left(\exp(i\beta {\bar{Q}}_{\text{matter}}) \otimes \exp(i\beta Q_{\text{ancilla}})\right)W={\bf{1}} \otimes \exp(i\beta Q_{\text{ancilla}}).
\end{align}
Since $Q_{\text{ancilla}}$ is on-site (eq.~\ref{eq:defQancilla}), this means that we have made $Q$ on-site by conjugating by the constant-depth circuit $W$.  We give some intuition for the disentangler $W$ and the necessity of ancillas in the next sub-section.

\subsection{Continuum intuition}
The disentangler $W$ may seem somewhat mysterious.  However, there is a continuum intuition for what it is doing (already articulated in ref. \cite{shirley2025}): it is performing a spatially varying $U(1)$ rotation on the `matter' degrees of freedom, controlled by the ancilla rotors $\theta_j$.  In other words, at site $j$ it performs a rotation on the matter fields by angle $\theta_j$, and this angle varies as a function of position $j$.  

Since the rotation of the matter degrees of freedom is not on-site, it is not obvious exactly how to spatially modulate it, and in fact this is where the anomaly-free condition comes in, as discussed below.  From this picture, it is clear why $W$ disentangles the symmetry: when we conjugate the trivial symmetry ${\bf{1}}\otimes \exp(i\beta Q_{\text{ancilla}})$ by $W$, i.e. take the operator $W \left({\bf{1}}\otimes \exp(i\beta Q_{\text{ancilla}})\right) W^\dagger$, what we are doing is: (1) rotating the matter fields backwards by the spatially dependent angles $-\theta_j$ (this is $W^\dagger$), (2) rotating the ancillas $\theta_j \rightarrow \theta_j+\beta$ (this is ${\bf{1}}\otimes \exp(i\beta Q_{\text{ancilla}})$), and finally (3) rotating the matter fields back by the spatially dependent angles $\theta_j+\beta$ (this is $W$).  The net effect is to rotate the ancillas by $\beta$ and to perform the desired global (not-on-site) rotation by $\beta$ on the matter degrees of freedom.

One may wonder how exactly the anomaly-free condition is used in this continuum intuition.  The point is that this spatially-dependent rotation of the matter fields must satisfy the group rules (i.e. rotating by $\{\theta_j\}$ and then $\{\theta_j'\}$ should be the same as rotating by $\{\theta_j+\theta_j'\}$).  This is not possible to do if the symmetry has an anomaly.  One way to see what goes wrong is to try to do the spatially-dependent rotation by
\begin{align}
\exp\left(i\sum_j \theta_jQ_{j-1,j}\right).
\end{align}
Let us examine the simple situation where $\theta_j$ is a non-zero constant $\theta$ on some interval of sites $I$, and is zero outside of it.  Then in particular we better have that
\begin{align}
\exp\left(\sum_{j\in I} 2\pi i Q_{j-1,j}\right) = \bf{1}.
\end{align}
But the self-anomaly is precisely the obstruction to this, because this operator pumps charge from one endpoint of $I$ to the other, and hence cannot be the identity.

\subsection{Hamiltonian for 3450 Bosons}

We unpack the general construction above in the case of 3450 bosons. This is a bosonized description of the 3450 chiral fermions consisting of two left-handed Weyls $\psi_L^\alpha$ of charge $3,4$ and two right-handed Weyls $\psi_R^\alpha$ of charge $5,0$. These satisfy anomaly-vanishing because $3,4,5$ is a Pythagorean triple:
\[3^2+4^2-5^2-0^2=0.\]
It is one of the simplest possible solutions to the anomaly-vanishing equation for chiral fermions.

The closely related theory of 3450 bosons has two compact bosons $\phi^\alpha$ with vector and axial charges $Q_{V^\alpha}$ and $Q_{A^\alpha}$. In terms of these charges, the desired $L/R$ gauge charges of the fermions can be expressed
\[(3,5)=4(1,1)-(1,-1)\\
(4,0)=2(1,1)+2(1,-1)\]
so we are interested in the gauge charge
\[\label{eqn3450gauge}Q_{\rm gauge}=4 Q_{V^1}-Q_{A^1}+2Q_{V^2}+2Q_{A^2}.\]

On the lattice we can thus have two species of rotors $\phi^\alpha$, each initially decoupled and obeying the Hamiltonian $H_{\rm Luttinger} = C H_{\rm Luttinger}^{\rm Vil} C^{-1}$ in \eqref{eqntensorluttingerham}, obtained by disentangling the Villain Luttinger Hamiltonian via $C$ in \eqref{eqnvildisentangler}. At low energies, this describes the $c=2$ theory of two compact bosons with the axial and vector charges above. Their individual marginal parameters are determined by $J_0/U_0$ in \eqref{eq:Luttinger}.

This Hamiltonian obeys all four symmetries $Q_{V^\alpha}=CQ_{V^\alpha}^{\rm Vil}C^{-1}$, $Q_{A^\alpha}=CQ_{A^\alpha}^{\rm Vil}C^{-1}$ defined in equations \eqref{eqnvectorvillain} and \eqref{eqnaxialvillain} respectively. The specific combination in \eqref{eqn3450gauge} above is anomaly-free, and therefore there is a disentangler $W$ constructed in \eqref{eq:defW}. To apply this disentangler, we add new rotors $\theta$, with the trivial Hamiltonian
\[H_0 = - \frac{d^2}{d\theta^2},\]
which has a product state ground state. Let $Q_{\rm ancilla} = \frac{i}{2\pi} \sum_j \frac{d}{d\theta_j}$. $H_0$ enjoys this symmetry and therefore $Q_{\rm gauge} + Q_{\rm ancilla}$ is a symmetry generator. The disentangler lets us write
\[Q_{\rm gauge} + Q_{\rm ancilla} = \sum_j W^\dagger q_j W\]
where $q_j$ are a set of commuting local operators with integer spectrum. We can proceed to introduce a $U(1)$ gauge field $E_{j+1/2},A_{j+1/2}$ living on the half-integer sites, together with the Gauss law
\[E_{j+1/2} - E_{j-1/2} = W^\dagger q_j W.\]
We are then free to add the usual gauge-invariant Maxwell Hamiltonian $\frac{1}{e^2}E_{j+1/2}^2$. This completes the construction of the 3450 bosons.

\section{Disentangling Chiral Symmetries in 3+1D}\label{sec:3+1d}

In this section we work on a branched simplicial decomposition of a spatial $3$-manifold $M^3$.  We first review the construction of vector and axial vector symmetries in the bosonic setting, where the Hilbert space consists of rotors $\phi \in C^0(M^3, \RR)$ \cite{fidkowski2025noninvertiblebosonicchiralsymmetry}.  Although $\phi$ is real-valued, the Hilbert space consists of wavefunctions invariant under individual integer shifts of $\phi$, so it in fact describes rotors.  A bosonic field theory saturating the mixed anomaly between these $U(1)_A$ and $U(1)_V$ symmetries was proposed in ref. \cite{fidkowski2025noninvertiblebosonicchiralsymmetry}.  We then show how to disentangle anomaly-free stacks of such theories.  Finally, we show how to modify this construction to include fermionic degrees of freedom in addition to the rotors, such that the anomaly is the same as that of the theory of four left-handed Weyl fermions with $U(1)_V$ and $U(1)_A$ charge assignments given in eq.~\ref{eqndiraccharges}.  We also generalize the disentangler to this fermionic setting, in particular allowing us to disentangle a theory matching the hypercharge assignments of one generation of the Standard Model, eq.~\ref{eqnSMhypercharges}.

\subsection{Bosonic axial symmetry review} \label{sec:bos_review}

In 3+1D, the $U(1)_A$ axial symmetry for a single rotor is diagonal in the rotor basis, and may be conveniently expressed as
\[\exp\left( i\beta \int_{M^3} \rho_A\right)=\exp \left(i \beta \int_{M^3} (\lr{d\phi}-d\phi) \cup d\lr{d\phi} \right),\qquad e^{i\beta} \in U(1)_A.\]
The local generator $(\lr{d\phi}-d\phi)\cup d\lr{d\phi}$ is invariant under local shifts $\phi \mapsto \phi+m$, $m \in C^0(M^3, \Z)$, since 
\[d\lr{d\phi} \mapsto d\lr{d\phi+dm} = d\lr{d\phi} + d^2m = d\lr{d\phi}.\]
Thus, this expression can be regarded as a finite-depth circuit upon expanding the exponential as a product:
\[\exp\left( i\beta \int_{M^3} \rho_A\right) = \prod_{\Delta^3 \subset M^3}\exp\left( i \beta \int_{\Delta^3} (\lr{d\phi}-d\phi) \cup d\lr{d\phi}  \right)\]
The individual terms are associated to tetrahedra $\Delta^3 \subset M^3$ and commute with one another. They can thus be staggered to obtain a constant depth circuit. It is also $2\pi$-periodic in $\beta$ on a closed manifold $M^3$. It can even be made $2\pi$-periodic on open manifolds by adding a counterterm:
\[\exp\left( i\beta \int_{M^3} \tilde \rho_A\right) = \exp \left(i \beta \int_{M^3} \rho_A + (d\phi - d\lr{\phi}) \cup d\lr{d\phi}\right) \\
= \exp \left( i \beta \int_{M^3} \rho_A + i \beta \int_{\partial M^3} (\phi-\lr{\phi}) \cup d\lr{d\phi}\right).\]
The counterterm is also locally shift invariant, so the local density $\tilde \rho_A$ is still well-defined and locally commuting, so we get a finite-depth circuit. Moreover, it does not change the symmetry action on a closed $M^3$, but now the combination is $2\pi$-periodic. In fact the local density $\tilde \rho_A$ is integer-valued. Thus, this $U(1)_A$ symmetry defined above is \emph{anomaly-free}.

We can also introduce a $U(1)_V$ symmetry, acting by shifts
\[\phi \mapsto \phi + \alpha, \qquad e^{i\alpha} \in U(1)_V.\]
This symmetry is on-site and thus anomaly-free as well. However, there is a \emph{mixed anomaly} between $U(1)_V$ and $U(1)_A$. Indeed, $U(1)_V$ preserves the unquantized local charge $\rho_A$, but not the quantized local charge $\tilde \rho_A$, so there is a tension between commutation and periodicity. More precisely, we may apply a $\beta = 2\pi$ axial rotation to a region $N^3$ with boundary, using the original definition with $\rho_A$. This produces a boundary term
\[\label{eqnbosonichallpump}\exp \left(-2\pi i \int_{\partial N^3} \phi \cup d[d\phi]\right).\]
This boundary term is an SPT entangler for a $U(1)_V$ SPT with 2+1D Chern-Simons term
\[\frac{1}{2\pi} A_V \wedge dA_V,\]
see \cite{demarco2021commutingprojectormodelnonzero}. As with the 1+1D charge pump diagnostic for the chiral anomaly in Section \ref{subsec1drotormodel}, this indicates a mixed $U(1)_V \times U(1)_A$ anomaly corresponding to the 4+1D Chern-Simons term
\[\label{eqnbosonicchiralanom4d}\frac{1}{4\pi^2} A_A \wedge dA_V \wedge dA_V.\]
There is one other possible mixed anomaly, of type $A_V (dA_A)^2$. For this type, a truncated $U(1)_V$ rotation would need to act as a $U(1)_A$ SPT pump. However, since $U(1)_V$ is on-site, we can truncate it in a $2\pi$-periodic manner, yielding no pump. Therefore, \eqref{eqnbosonicchiralanom4d} is the entire anomaly.

If we have several species of rotors, with axial charges $q_A^\alpha$ and vector charges $q_V^\alpha$, the anomaly for the \emph{diagonal symmetry} is
\[\frac{1}{4\pi^2} \bigg(\sum_\alpha q_A^\alpha q_V^\alpha q_V^\alpha \bigg) A_A \wedge dA_V \wedge dA_V.\]
The diagonal symmetry is thus anomaly-free if and only if
\begin{align} \label{eqn:3danomaly_cancellation}
\sum_\alpha q_A^\alpha q_V^\alpha q_V^\alpha = 0.
\end{align}
In the next section we show that under this condition, the diagonal symmetry can indeed be disentangled with the introduction of ancillas. It can then be gauged in the usual way.

\subsection{The bosonic disentangler}
We now assume we have a stack of the $3+1$ dimensional systems introduced above, labeled by $\alpha$, with a $U(1)$ generator $Q = \sum_{\alpha} (q_V^\alpha Q_V +q_A^\alpha Q_A)$.  As in the $1+1$ dimensional case, we introduce ancilla rotors $\theta_j$, and define our disentangler $W$ by:
\begin{align}
W|\phi^\alpha, \theta\rangle = W_{\phi^{\alpha}, \theta }\,|\phi^{\alpha} + q^{\alpha}_V \theta, \theta\rangle
\end{align}
where $W_{\phi^{\alpha}, \theta}$ is a phase factor.  We claim that the following choice of $W_{\phi^{\alpha}, \theta}$ disentangles $Q$:
\begin{align}
W_{\phi^{\alpha}, \theta} = \exp\left[-i \int_{M^3} \omega(\phi^\alpha,\theta) \right],
\end{align}
with
\[\omega(\phi^\alpha,\theta)= \sum_\alpha\Bigg(\theta \Big(q_A^\alpha \lr{d\phi^\alpha+q_V^\alpha d\theta}-d\lr{q_A^\alpha \phi^\alpha}-q_A^\alpha q_V^\alpha d\lr{\theta} \Big)\cup d\lr{d\phi^\alpha + q_V^\alpha d\theta}
\\+ q_A^\alpha q_V^\alpha d\theta \cup \Big(d\lr{\theta}-\lr{d\theta} \Big) \cup \Big(\lr{d\phi^\alpha + q^\alpha_V d\theta}-d\lr{\phi^\alpha} \Big)\Bigg).\]
The first term in $\omega(\phi^\alpha,\theta)$ may be easily checked to be shift invariant, up to integers, for both $\theta$ and $\phi^\alpha$. The second term is also clearly shift invariant for $\phi^\alpha$. For $\theta \mapsto \theta + n$, it transforms by
\[\sum_\alpha q_A^\alpha q_V^\alpha q_V^\alpha d\theta \cup \Big(d\lr{\theta}-\lr{d\theta} \Big) \cup dn,\]
which vanishes when we sum over $\alpha$ by the anomaly vanishing equation
\[\sum_\alpha q_A^\alpha q_V^\alpha q_V^\alpha=0.\]

Using the same argument as in $1{+1}$D, showing that our disentangler indeed disentangles the symmetry is equivalent to showing that, on a closed manifold,
\begin{align}
\Delta_\beta \omega &\equiv\omega(\phi^\alpha +q_V^\alpha \theta,\theta+\beta)-\omega(\phi^\alpha +q_V^\alpha \theta,\theta) \\ &=\beta \sum_\alpha q_A^\alpha \rho_A.
\end{align}
To see that this is the case, let us rearrange the terms to
\[\omega(\phi^\alpha,\theta) = \sum_\alpha \Bigg(\theta \Big(q^\alpha_A \lr{d\phi^\alpha+q_V^\alpha d\theta}-d\lr{q_A^\alpha \phi^\alpha} \Big) \cup d\lr{d\phi^\alpha + q_V^\alpha d\theta} \\
- q_A^\alpha q_V^\alpha \theta d\lr{\theta} \cup d\lr{d\phi^\alpha + q_V^\alpha d\theta} \\
+  q_A^\alpha q_V^\alpha d\theta \cup \Big(d\lr{\theta}-\lr{d\theta} \Big) \cup \Big(\lr{d\phi^\alpha + q^\alpha_V d\theta}-d\lr{\phi^\alpha} \Big)\Bigg).\]
Call the first term $\omega_0$ and the rest of the terms $\omega_1$, so $\omega = \omega_0 + \omega_1$. The first term $\omega_0$ can be checked to be a disentangler, i.e. it alone contributes the correct quantity to $\Delta \omega$, namely
\[\Delta_\beta \omega_0 = \sum_\alpha \beta q_A^\alpha \lr{d\phi^\alpha} \cup d\lr{d\phi^\alpha} + d(\cdots) = \beta\sum_\alpha q^\alpha_A \rho_A + d(\cdots)\] 
which defines an axial rotation by $\beta$ on a closed 3-manifold. Thus, we just want to show the other two terms contribute a vanishing quantity $\Delta_\beta \omega_1 = d(\cdots)$. We compute
\[\Delta_\beta \omega_1 = \sum_\alpha \Bigg( -q_A^\alpha q_V^\alpha (\theta+\beta) d\lr{\theta+\beta} \cup d\lr{d\phi^\alpha} +q_A^\alpha q_V^\alpha (\theta) d\lr{\theta} \cup d\lr{d\phi^\alpha} \\
+ q_A^\alpha q_V^\alpha d\theta \cup \bigg(d\lr{\theta+\beta}-d\lr{\theta} \bigg) \cup \bigg(\lr{d\phi^\alpha}-d\lr{\phi^\alpha - q^\alpha_V \theta}\bigg) \Bigg)\]
\[\label{eqndisentanglestep2}= \sum_\alpha \Bigg( -q_A^\alpha q_V^\alpha \beta \bigg(d\lr{\theta+\beta} \bigg) \cup d\lr{d\phi^\alpha}
\\-q_A^\alpha q_V^\alpha \theta \bigg(d\lr{\theta+\beta}-d\lr{\theta} \bigg) \cup d\lr{d\phi^\alpha} \\
+ q_A^\alpha q_V^\alpha d\theta \cup \bigg(d\lr{\theta+\beta}-d\lr{\theta} \bigg) \cup \bigg(\lr{d\phi^\alpha}-d\lr{\phi^\alpha - q^\alpha_V \theta}\bigg) \Bigg).\]
The first term in \eqref{eqndisentanglestep2} is a total derivative, so it will not contribute anything on a closed manifold. It turns out the rest of the terms are a total derivative as well, namely:
\[d\Bigg(q_A^\alpha q_V^\alpha \theta \cup \bigg(d\lr{\theta+\beta}-d\lr{\theta} \bigg) \cup \bigg(\lr{d\phi^\alpha}-d\lr{\phi^\alpha - q^\alpha_V \theta} \bigg) \Bigg).\]
Thus, $\Delta \omega_1 = d(\cdots)$, as required.

\subsection{Fermionic axial symmetry}

Above we have considered symmetry disentanglers for chiral symmetries of bosons in $3{+}1$D. Now we would like to extend these constructions to fermions, and make the connection to the target IR theory of four left-handed Weyl fermions with charges in eq.~\ref{eqndiraccharges}. We will have to augment the Hilbert space to include fermionic degrees of freedom in addition to the bosonic rotors we already have. We will also have to modify our definition of $U(1)_A$ to act on these fermions. $U(1)_V$ will not be modified.

More precisely, the target IR theory has a symmetry $U(1)_V \times U(1)_A$ with the charges
\begin{align}\begin{matrix}
    U(1)_V & +1 & -1 & 0 & 0 \\
    U(1)_A & +1 & +1 & -1 & -1
\end{matrix}\end{align}
In particular a $\pi$ rotation in $U(1)_A$ corresponds to fermion parity. The anomaly is computed in \eqref{eqnappanom} and can be expressed as a Chern-Simons term
\[\int_{X^5} A_A (dA_V/2\pi)^2\]
where $A_A$ and $A_V$ are the background gauge fields for $U(1)_A$ and $U(1)_V$ respectively.\footnote{The former should be considered a Spin$^c$ structure.} From this Chern-Simons term we can conclude that a $2\pi$ axial rotation in a finite region produces a $U(1)_V$ SPT on its boundary with the topological response
\[\label{eqnfermionpumpedsptfullperiod}\int_{Y^3} \frac{1}{2\pi} A_V dA_V\]
with a Hall conductivity of 2. We will reproduce this pump in the lattice model.

The pump can be seen directly in terms of the free fermion QFT, giving another method of computing its anomaly \cite{debray2024longexactsequencesymmetry}. Consider breaking the $U(1)_A$ symmetry to $(-1)^F$ while preserving $U(1)_V$ by turning on a Dirac mass pairing the first and second Weyls (and separately the third and fourth Weyls) above. If we perform a $\pi$ rotation in $U(1)_A$ in a region, the complex phase of this Dirac mass winds by $2\pi$ for the first pair and $-2\pi$ for the second pair. This produces at the boundary of the region a $c_1=1$ Chern insulator\footnote{This pumping is well-known, and can be understood via the boundary of this Chern insulator. This boundary can be considered a vortex in the Dirac mass, where each pair of Weyls will contribute a single $1{+}1$D Weyl along the vortex of opposite chirality \cite{CallanHarvey_vortex}. See also \cite{Fidkowski_Xu, fidkowski2025noninvertiblebosonicchiralsymmetry}.} of the first pair, which carries unit $U(1)_V$ charge, and a $c_1 = -1$ Chern insulator of the second pair, which is $U(1)_V$ neutral. This stack of Chern insulators corresponds to half of the SPT in eq.~\ref{eqnfermionpumpedsptfullperiod} since we have only done a $\pi$ rotation. As a by-product of our construction, we will obtain a new commuting-projector Hamiltonian for this $2{+}1$D SPT.

\begin{figure}[H]
    \centering
    \includegraphics[width=0.62\linewidth]{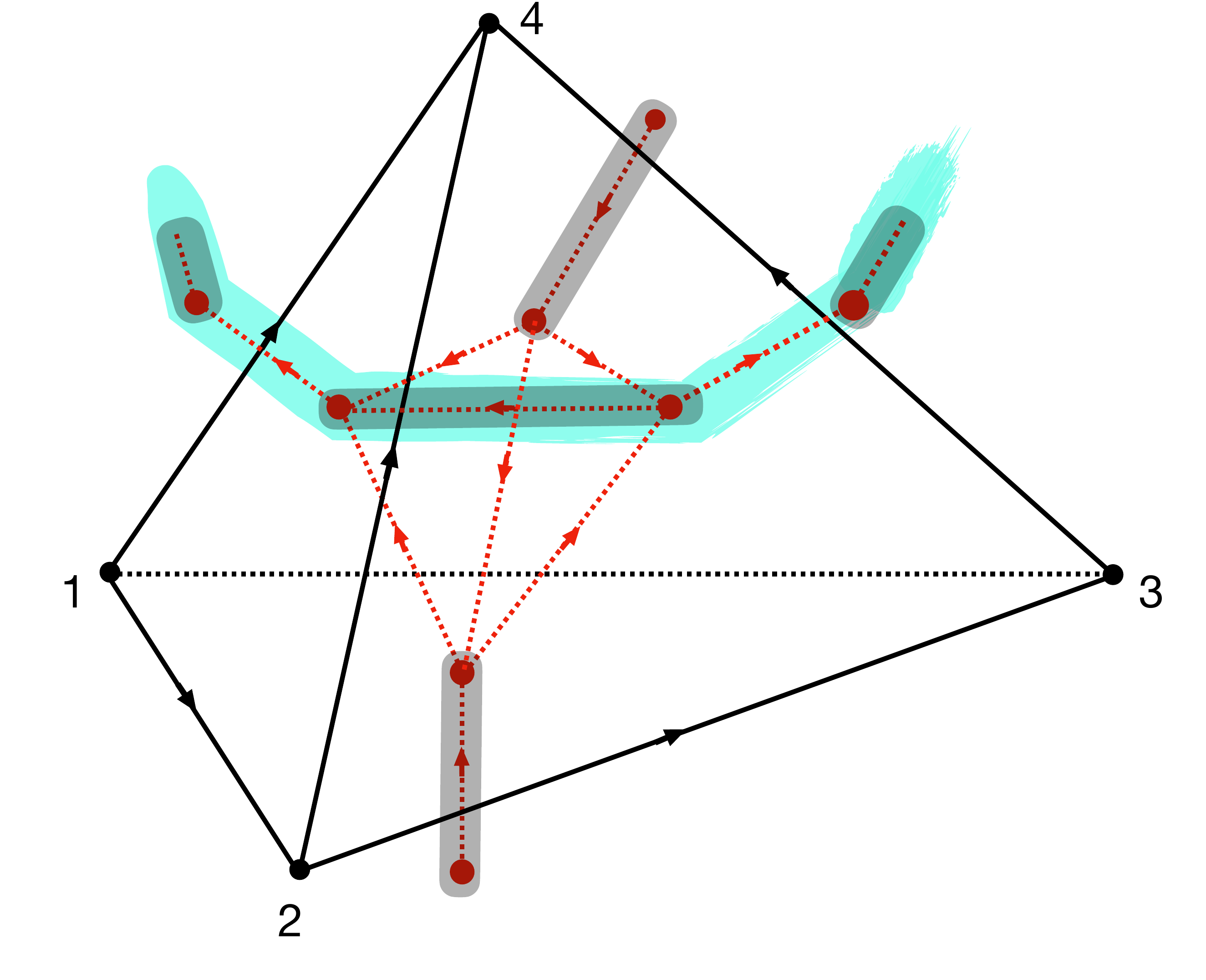}
    \captionsetup{width=\linewidth}
    \caption{
    Lattice model of rotors and fermions, with the fermionic decoration adapted from \cite{Wang_2018, Wang_2020}.  The black tetrahedron shown is part of a simplicial decomposition of the spatial $3$-manifold.  This simplicial decomposition has a branching structure, which we may take to be induced from a global ordering of the black vertices: orientations on the black edges point from smaller to larger numbers.  The rotors are located at the black vertices.  The red dots represent Majorana fermions.  Each physical fermionic degree of freedom is represented by a pair of such Majorana fermions straddling a triangular face of the black tetrahedron.  In the above figure we show $8$ Majorana fermions, corresponding to $4$ physical fermions associated to the $4$ faces of the black tetrahedron.  The dotted red lines form the `resolved dual lattice' in the terminology of \cite{Wang_2018, Wang_2020}, who show how to define a so-called Kasteleyn orientation on it.  The fermionic part of the $U(1)_A$ generator (eq.~\ref{eq:defK}) is defined using a fermionic bilinear where the Majorana fermions are paired according to a pairing $p_\phi$ controlled by the rotor configuration $\{\phi\}$.  This pairing just decorates Kitaev chains on odd vorticity loops.  More precisely, the Poincare duals of the black triangular faces carrying odd vorticity $d\lr{d\phi}$ form closed loops (blue) in which the Majorana fermions are paired within the black tetrahedra; the remaining Majorana fermions are paired across the black triangular faces.  In the figure above, the vortex penetrates the faces $124$ and $234$.}
    \label{fig:4simplex}
\end{figure}

\begin{figure}[H]
    \centering
    \includegraphics[width=0.52\linewidth]{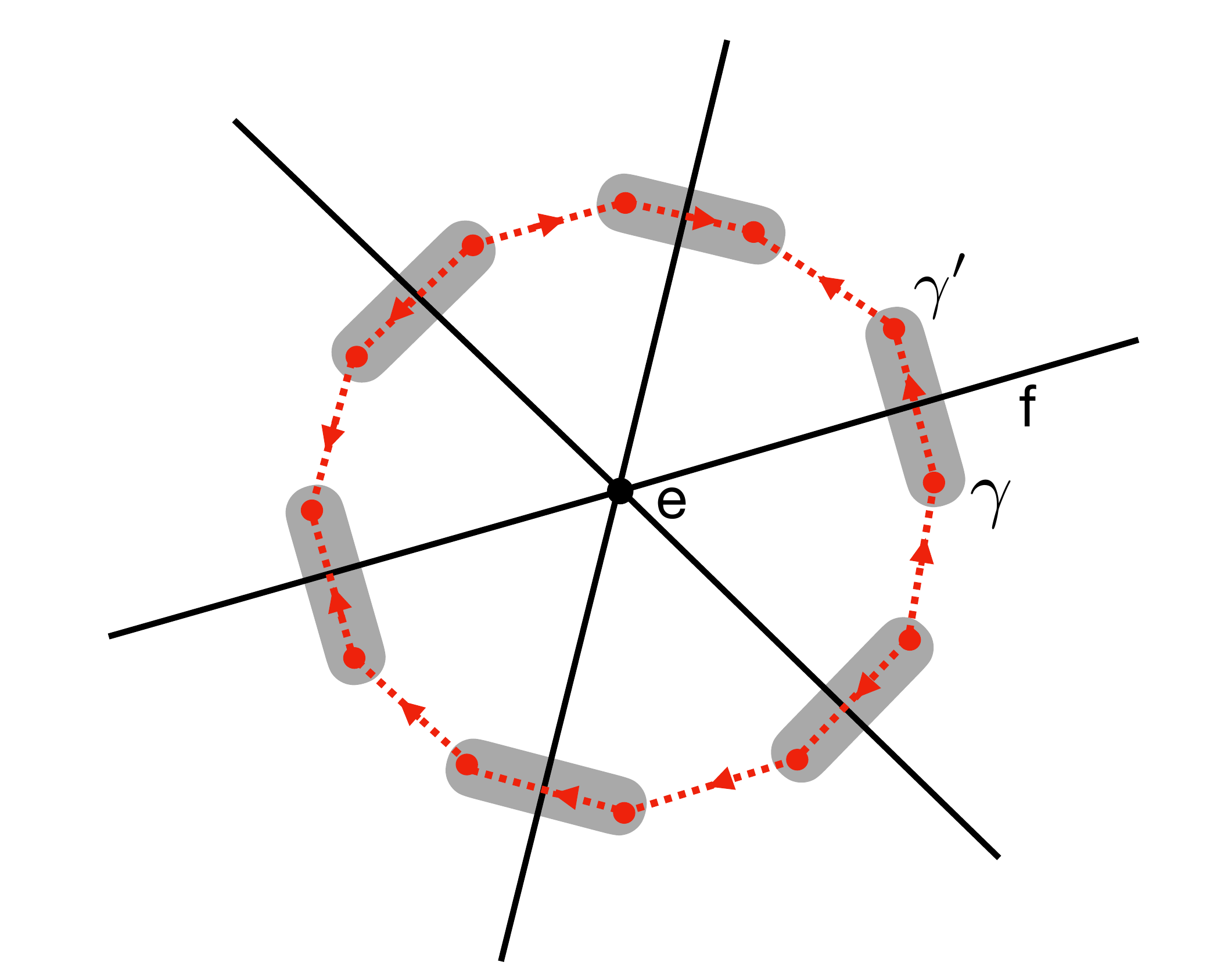}
    \captionsetup{width=\linewidth}
    \caption{Short loop of resolved dual edges (dotted red lines) surrounding the edge $e$ on the resolved dual lattice.  For clarity we have projected from $3$ dimensions to $2$ dimensions, so the central point represents an edge $e$ oriented perpendicular to the page and the black lines represent faces $f$ that contain $e$.  The short loop is Kasteleyn oriented, because an odd number of its edges are oriented clockwise, or counter-clockwise (since there are always an even number of resolved dual edges in any such short loop, this property is independent of the choice of clockwise or counter-clockwise orientation along the loop).  The red dots represent Majorana fermions, and the gray ovals represent the `trivial' pairing $p_0$, with each such pair defining a complex fermion.  The Kasteleyn property implies that if we instead chose the complementary pairing $p_1$ of the Majorana fermions, i.e. formed a Kitaev chain along the loop,  then $\prod_{\langle \gamma \gamma' \rangle \in p_0} i\gamma\gamma' = \prod_{\langle \gamma \gamma' \rangle \in p_1} i\gamma\gamma'$, where by convention the orientations always point from $\gamma$ to $\gamma'$. }
    \label{fig:Kateleyn_loop}
\end{figure}

Our geometry will be as follows.  As in the bosonic case, we have a branched simplicial decomposition of a three-manifold $M^3$, closed for now, with the $\phi$ degrees of freedom living at the vertices (later we will take several copies $\phi^\alpha$, and also include ancilla $\theta$ degrees of freedom living at these vertices).  We add a single complex fermion to each 2-simplex (face), and view this fermion as a pair of Majorana fermions, one on either side of the face.  These Majorana fermions form the vertices of the `resolved dual lattice' in the terminology of Ref. \cite{Wang_2020} (figure \ref{fig:4simplex}).  The edges of this resolved dual lattice consist of the original pairing of the Majoranas across each face, as well as all six pairings of the four Majoranas within each $3$-simplex.  Refs. \cite{Wang_2018, Wang_2020} show how to construct a Kasteleyn orientation of the edges of the resolved dual lattice.  This Kasteleyn orientation is characterized by the following property.  Given any edge $e$ of the original lattice, consider the set of resolved dual lattice edges which are dual to $2$-simplices that contain $e$, together with (the same number of) resolved dual edges that turn this set into a loop $l_e$.  Essentially this is the shortest loop in the resolved dual lattice that encircles $e$, and is given by `cutting the corners' off the $2$-simplex dual to $e$ - see figure \ref{fig:Kateleyn_loop}.  Now pick an arbitrary orientation along $l_e$, and count how many resolved dual edges in $e$ are consistent with this orientation.  If this number is odd, we say the loop is `Kasteleyn-oriented'.  The Kasteleyn property is that all such short loops $l_e$ are Kasteleyn-oriented \cite{Wang_2018}.

The Kasteleyn orientation constructed in Refs. \cite{Wang_2018, Wang_2020} actually has a more general property.  Namely, we can extend the notion of being Kasteleyn-oriented to a general closed loop $l$, Poincare dual to some $\Z_2$-valued $2$-form $n_2$ on the original lattice, by again defining a corresponding loop on the resolved dual lattice by `cutting the corners'.\footnote{There is an ambiguity in how to cut the corners when all $4$ faces of a tetrahedron have $n_2 \neq 0$; Ref. \cite{Wang_2018} picks a specific convention for resolving this ambiguity.}  Then $l$ is Kasteleyn-oriented if and only if its self-linking number $\int_{M^3} m_1 \cup n_2$ is even, where $dm_1 = n_2$ \cite{Wang_private_communication}.

Let us now fix a configuration of the rotors $\phi$, and specialize to $n_2 \equiv d\lr{d\phi} \text{ mod } 2$, $m_1 \equiv \lr{d\phi} \text{ mod } 2$.  Then $n_2$ just represents the faces penetrated by odd vorticity of $\phi$, and the Poincare dual of $n_2$ is the set of vortex loops of odd vorticity.  Given such $n_2$ we define a pairing of Majorana fermions as follows.  Away from the odd vortices, i.e. on the triangles where $n_2 = 0$, we pair the two Majoranas corresponding to each such triangle with each other.  Along the odd vortices, we perform the complementary pairing, so that the vortices are decorated by Kitaev chains.  This pairing is described in detail in Refs. \cite{Wang_2018, Wang_2020}; let us denote it as $p_\phi$, with the subscript emphasizing that it depends on the configuration of $\phi$ fields.  We use the convention that for every pair $\langle\gamma, \gamma'\rangle \in p_\phi$, the order of $\gamma$ and $\gamma'$ is dictated by the orientation of the corresponding resolved dual edge.  We then define the operator
\begin{align} \label{eq:defK}
K =  \int {\cal{D}}\phi \frac{1}{2}\left(1- \sum_{\langle\gamma, \gamma'\rangle \in p_\phi} i \gamma \gamma' \right)|\phi\rangle \langle \phi|.
\end{align}
Note that this is a free-fermion operator controlled by the state of the rotors $\phi$.  Since $\exp(\pi i (1-i \gamma \gamma')/2) = i\gamma\gamma'$, we have that on a closed manifold
\begin{align} \label{eq:ferm_exp}
\exp(i\pi K) = \prod_{\langle\gamma, \gamma'\rangle \in p_\phi} i\gamma\gamma'=(-1)^{\int_{M^3} \lr{d\phi} \cup d\lr{d\phi}} \cdot (-1)^F
\end{align}
where we define the fermion parity
\begin{align}
(-1)^F = \prod_{\langle\gamma, \gamma'\rangle \in p_{0}} i\gamma\gamma'
\end{align}
and $p_{0}$ is the trivial pairing associated to a constant $\phi$ configuration (no vortices).  Eq.~\ref{eq:ferm_exp} follows from the Kasteleyn property of the orientation, and in fact just reduces to the familiar fact that in a one dimensional fermionic lattice system on a ring, the product of $i \gamma \gamma'$ over the non-trivial Kitaev chain pairing is equal to the product over the trivial pairing precisely when the orientation satisfies the Kasteleyn property \cite{Tarantino} - see figure \ref{fig:Kateleyn_loop}.

We now define the fermionic axial symmetry generator
\begin{align} \label{eq:fermionic_QA}
Q_A^{\text{f}} = K+\int{\cal{D}}\phi\int_{M^3} ([d\phi]-d\phi) \cup d[d\phi]\,|\phi\rangle\langle\phi|.
\end{align}
We note, using eq.~\ref{eq:ferm_exp}, that on a closed manifold $M^3$, $\exp(i\pi Q_A^{\text{f}}) = (-1)^F$. Furthermore, when truncated to a finite region, the bosonic piece of $\exp(i\beta Q_A^{\text{f}})$ pumps a 2+1D SPT state with Hall conductance of $2$ when $\beta = 2\pi$, corresponding to the Chern-Simons term in eq.~\ref{eqnfermionpumpedsptfullperiod} (see also eq.~\ref{eqnbosonichallpump}). Thus, this lattice symmetry realizes the anomaly \eqref{eqnappanom} of the $4$-Weyl-fermion theory in eq.~\ref{eqndiraccharges}.

Let us now comment on the $\pi$ pump we described above in the free fermion QFT.  In particular, we can regard $\exp(i\pi Q_A^{\text{f}})$ as a pump of a 2+1D SPT with Hall conductance $1$, since it acts as fermion parity in the bulk, which is on-site. By standard arguments \cite{debray2024longexactsequencesymmetry}, what is pumped may be regarded as a $2+1$D SPT state of fermions protected by $U(1)_V \times (-1)^F$ symmetry. From the classification of SPT phases, the only such SPT with Hall conductance $1$ is (adiabatically connected to) a stack of a Chern number $+1$ Chern insulator of $U(1)_V$-charged fermions, together with a Chern number $-1$ Chern insulator of neutral fermions.  This SPT pump can also be seen directly in the free fermion picture of the $4$ Weyl fermion theory in eq.~\ref{eqndiraccharges}.  Indeed, the $4$ Weyl fermions can be viewed as $2$ Dirac fermions, one charged under $U(1)_V$ and the other neutral.  The $U(1)_A$ acts in opposite ways in the $2$ Dirac fermions, and hence pumps opposite Chern number Chern insulators, one $U(1)_V$-charged and one neutral.

We can also analyze more closely the wavefunction of the boundary SPT state pumped by $\exp(i\pi Q_A^{\text{f}})$, and confirm that it has the universal properties we expect.  By the same arguments as given in \cite{demarco2021commutingprojectormodelnonzero}, this state has a commuting-projector parent Hamiltonian.\footnote{This Hamiltonian may be obtained by applying the truncated $\exp(i\pi Q_A^{\text{f}})$ to a trivial $U(1)_V \times (-1)^F$ symmetric commuting-projectorHamiltonian, leveraging the on-siteness of this symmetry. This produces a Hamiltonian which near the boundary of the truncation is non-trivial but away from that boundary is trivial. It can thus be truncated to obtain a symmetric 2+1D commuting-projectorHamiltonian.}  In particular, when $Q_A^{\text{f}}$ is truncated to a finite $3$ dimensional region $R$, the resulting truncated generator, denoted ${\bar{Q}}_A^{\text{f}}$ will have un-paired Majorana modes where the odd vortices meet the boundary.  Then $\exp(i\pi {\bar{Q}}_A^{\text{f}})$ is equal to a $\phi$ dependent phase times the fermion parity in $R$, times the product of these un-paired Majorana modes.  Thus the fermionic SPT state living on $\partial R$ must have odd fermion parity at these vortex cores.  This is consistent with the stack of Chern insulators picture, because if we think of an odd vortex as a single magnetic flux quantum of $U(1)_V$, we expect only the charged Chern insulator to respond to it, pulling in a single $U(1)_V$ charge (due to the $U(1)_V$ Hall conductance being $1$), as well as fermion parity.

\subsection{The fermionic disentangler}

As in the bosonic case, we now consider a stack of $N$ fermionic systems, labeled by a layer index $\alpha = 1,\ldots,N$, and also introduce ancilla rotors $\theta$ at the vertices.  We again define a diagonal $U(1)$ symmetry
\begin{align}
Q^{\text{f}}_{\text{matter}} = \sum_{\alpha} (q_V^\alpha Q_V +q_A^\alpha Q_A).
\end{align}
By the same arguments as in the bosonic case (section \ref{sec:bos_review}) the anomaly cancellation condition is
\begin{align} 
\sum_\alpha q_A^\alpha q_V^\alpha q_V^\alpha = 0.
\end{align}
Denote the bosonic disentangler by $W$.  Recall that it has the property that
\begin{align}
 W\left({\bf{1}} \otimes \exp(i\beta Q_{\text{ancilla}})\right)W^\dagger=\exp(i\beta Q_{\text{matter}}) \otimes \exp(i\beta Q_{\text{ancilla}}).
\end{align}
To construct a fermionic disentangler, we seek to dress $W$ with a fermionic piece $W'$ such that
\begin{align} \label{eq:ferm_disentang}
W W'&\left({\bf{1}} \otimes \exp(i\beta Q_{\text{ancilla}})\right)(W')^\dagger W^\dagger = \exp(i\beta Q^{\text{f}}_{\text{matter}}) \otimes \exp(i\beta Q_{\text{ancilla}}).
\end{align}
Specifically, we will define
\begin{align}
W' = \int {\cal{D}}\phi^\alpha \prod_\alpha \exp\left(\sum_{\langle \gamma,\gamma'\rangle \in p_{\phi^\alpha}} \theta_{\gamma,\gamma'} \cdot \gamma \gamma' \right)|\phi^\alpha + q_V^\alpha \theta\rangle \langle\phi^\alpha + q_V^\alpha \theta|.
\end{align}
Here $\theta_{\gamma,\gamma'}$ is just $\theta_j$ for some $j$ near the pair $\langle \gamma,\gamma'\rangle$ - we can just arbitrarily pick such an assignment $\langle \gamma,\gamma'\rangle \rightarrow j$ once and for all (also independently of $\alpha$).  

It is then easy to verify eq.~\ref{eq:ferm_disentang}.  Essentially, the point is that $(W')^\dagger$ performs a local rotation by $-\theta_j$ on the fermionic degrees of freedom, and $W'$ performs a rotation by $\theta_j+\beta$, resulting in a net constant rotation by $\beta$, i.e. $\exp(i\beta K)$, as desired.  So the fermionic $U(1)$ axial symmetry can be disentangled into one that acts just on the bosonic rotors $\theta_j$.

\section{In-flow point of view} \label{sec:in-flow}

Our constructions so far start in $D$ spacetime dimensions, using Hilbert spaces in which $G$ acts not-on-site, thereby encoding prescribed 't Hooft anomalies. In this section we describe a related formulation based on an anomaly in-flow picture, which more closely resembles the SMG scenario. The $D$-dimensional model with not-on-site symmetry is replaced by a boundary truncation, built using an SPT disentangler\footnote{This SPT entangler is built using a swindle construction from our symmetry disentanglers applied to pairs of $D$ dimensional models with trivially cancelling anomalies, see Appendix \ref{appcphfromdisentangler}.}, of a $(D{+1})$-dimensional SPT with a commuting projector Hamiltonian. In this scheme, realizing the chiral theory reduces to constructing a symmetric trivial gapped interface between two SPTs (free-fermion and commuting-projector) which are expected to lie in the same phase.

\begin{figure}
    \centering
    \includegraphics[width=10cm]{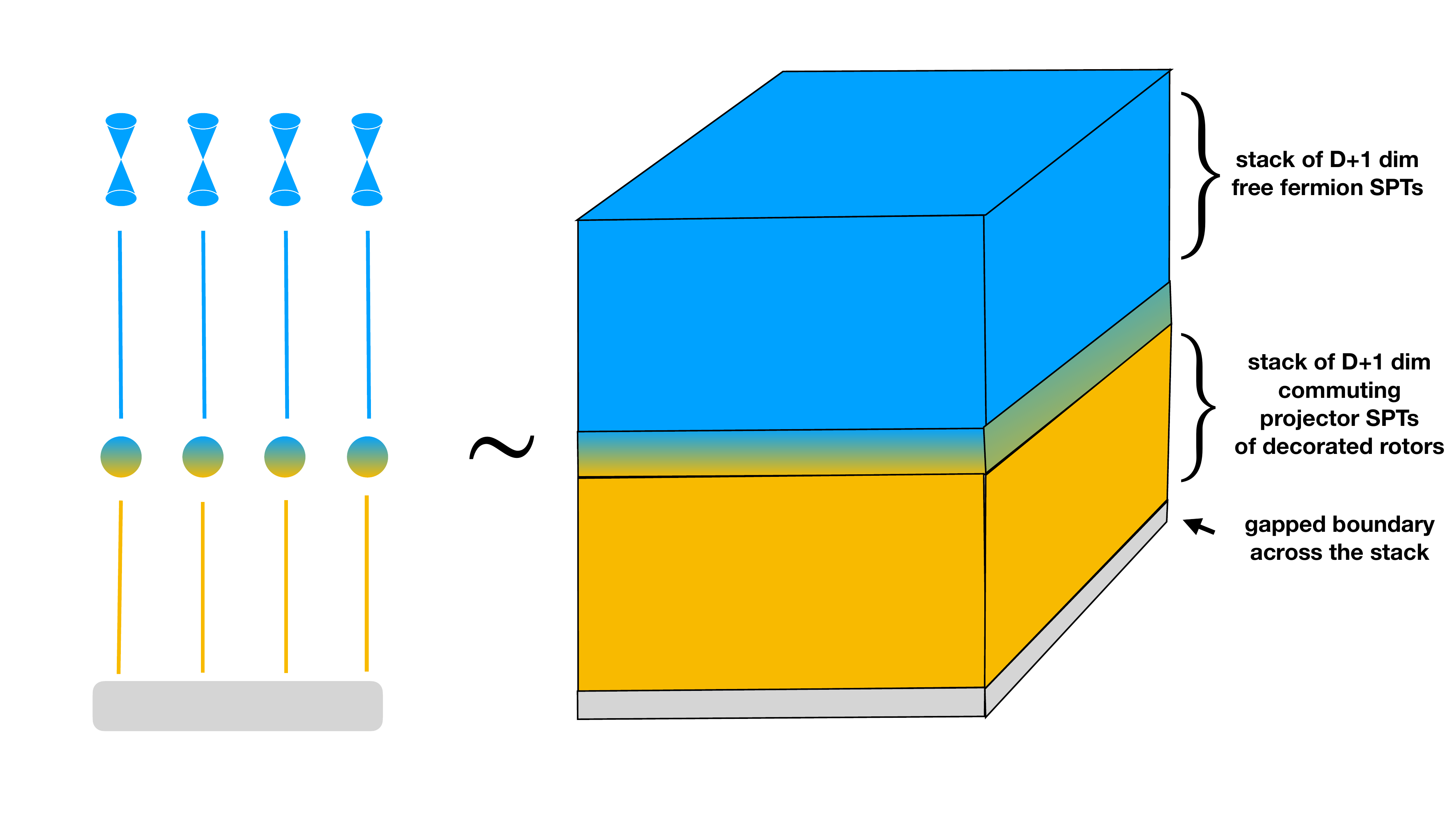}
    \caption{Variant of the $D>2$ construction from the anomaly in-flow perspective. A stack of $N$ $(D{+}1)$-dimensional free-fermion $G$-SPTs (blue slab) is coupled across an interface (blue-yellow region) to a stack of $(D{+}1)$-dimensional commuting-projector $G$-SPTs (yellow slab). This interface is assumed to be fully gapped, topologically trivial, and preserves the $G$ symmetry, while the top surface of the blue slab remains gapless. A symmetry disentanger (grey) supported near the bottom of the yellow slab, renders on-site an anomaly-free subgroup $G'\subset G^N$, enabling the bottom boundary to be trivially gapped by an on-site Hamiltonian without breaking $G'$. This construction provides a Hamiltonian realization of the SMG scenario for constructing chiral gauge theories.}

    \label{fig:3450-figureB}
\end{figure}

The construction we will now describe has three steps summarized in Figure \ref{fig:3450-figureB}. ($i$) Using the SPT disentangler, we construct an SPT with commuting-projector Hamiltonian $H_{CP}$ and on-site $G$ symmetry whose $D$-dimensional boundary truncation realizes the anomalous symmetry action. ($ii$) The upper surface of a $(D{+}1)$-dimensional slab of the $H_{CP}$ commuting-projector model is trivially gapped against the lower surface of a $(D{+}1)$-dimensional slab of a free-fermion $G$-SPT. ($iii$) For a stack of $N$ such slabs with canceling anomalies under $G'\subset G^N$, we apply a $G'$ symmetry disentanger near the bottom boundary to render $G'$ on-site and gap that boundary with an explicit on-site Hamiltonian.

How the first step is accomplished is described in Appendix \ref{appcphfromdisentangler}. There we review a standard construction which shows, given a possibly anomalous, not-on-site $D$ dimensional unitary representation of a symmetry group $G$, how to construct a $(D{+}1)$-dimensional commuting-projector Hamiltonian $H_{CP}$ with on-site $G$ symmetry.  Truncating $H_{CP}$ to a boundary reproduces the $D$-dimensional not-on-site, anomalous symmetry action.

The idea behind the second step is that we can replace each $D$-dimensional model in our previous construction by the top boundary of a $(D{+1})$-dimensional slab with commuting-projector Hamiltonian $H_{CP}$. We then place this commuting projector slab adjacent to a free-fermion $G$-SPT, and attempt to achieve a trivial gapped interface between the bottom surface of the free-fermion SPT and the top surface of the commuting projector slab while preserving the $G$ symmetry. This trivial interface should exist if the two models are representatives of the same SPT phase~\cite{czajka2025anomalieslatticehomotopyquantum}. This can plausibly be expected for example in the specific case of $G=U(1)_V \times U(1)_A$ symmetry studied in this work because the free-fermion and commuting-projector models have the same Chern-Simons terms for background $U(1)_V \times U(1)_A$ gauge fields;  continuum SPT phases are believed to be classified by this topological response~\cite{Kapustin_2015}, suggesting that these lattice models also lie in the same phase.

After gluing together the two SPT slabs, the $G$ symmetry action is on-site throughout the bulk. The remaining issue, to be addressed by the third step, is that the bottom surface of the commuting-projector stack still carries low-energy boundary degrees of freedom. Suppose, though, that for a stack of $N$ layers, the subgroup $G'\subset G^N$ is anomaly-free. In this case, using the $G'$ symmetry disentangler as described in Appendix \ref{appcphfromdisentangler}, we can construct an exactly solvable boundary Hamiltonian with $G'$ symmetry that acts across the stack and trivially gaps the bottom boundary. To achieve this, first a truncated SPT disentangler decouples the commuting projector bulk systems from their bottom boundaries, so that $G$ has not-on-site action on the bottom boundaries. Then the $G'$ disentangler applied across the stack renders $G'$ on-site on the bottom boundary. Once $G'$ is on-site, it can be trivially gapped by an on-site symmetric Hamiltonian-- for example, one can gap a $U(1)$ symmetric rotor with $H=J^2$, where $J$ is angular momentum.  Because everything here takes place in a zero-correlation-length setting, this procedure provides a well-controlled lattice realization of the SMG scenario. Only step ($ii$) is not fully rigorous, as the existence of the symmetric gapped interface between the free-fermion and commuting-project SPTs is a plausible but unproven conjecture, which we discuss further in \ref{subsec:gapped-surface}.

\subsection{Gapped surface between free fermion and commuting-projector SPT}
\label{subsec:gapped-surface}

In particular, the method above can be applied to the hypercharge sector of the Standard Model.  In that case $G = U(1)_V \times U(1)_A$, and the corresponding SPT has a free-fermion representative.  Namely, there is a 4+1D free-fermion SPT of $U(1)$ -- the 4+1D integer quantum Hall effect, in ``class A'' - that has a single left-handed Weyl on its boundary.  By stacking $4$ copies of this ``root'' SPT and considering the $U(1)_V \times U(1)_A \subset U(1)^4$ subgroup of this stack defined by \eqref{eqndiraccharges}, we get the desired free fermion SPT.  This SPT phase also has a commuting-projector Hamiltonian, according to the general construction in Appendix \ref{appcphfromdisentangler}.  The problem of realizing the hypercharge sector of the Standard Model is then reduced to constructing a gapped interface, trivial in the sense of having no topological order, between these free-fermion and commuting-projector Hamiltonians, which are putatively in the same SPT phase. We now discuss a strategy for constructing such an interface.

First, we consider $U(1)_V$-breaking, $U(1)_A$-symmetric states on the rotor model side where $\phi$ are in an arbitrary product state. This can be achieved by pairing the fermions into a trivial state according to the pairing $p_{\phi}$ used in defining the $U(1)_A$ charge, i.e. $i\gamma \gamma'=1$ for every $\langle \gamma \gamma' \rangle \in p_{\phi}$, which commutes with $K$ (see eq.~\ref{eq:defK}). This is a state whose $U(1)_A$ eigenvalue is equal to the (integer) self-linking number of the vortex configuration $d\lr{d\phi}$ (eq.~\ref{eq:fermionic_QA}).  Now suppose we want to (naively) construct a combination of these states which also has the $U(1)_V$ symmetry by allowing $\phi$ to fluctuate. For instance, we could add terms proportional to $-\frac{d^2}{d\phi^2}$ to the Hamiltonian, producing a trivial gapped ground state which is an equal superposition over all rotator states with the fermions paired according to $p_\phi$.  However, this ends up breaking the $U(1)_A$ symmetry, since these new terms fluctuate between states with different $U(1)_A$ eigenvalues by changing the vortex self-linking number.  Indeed, this is a necessary consequence of the mixed $U(1)_V \times U(1)_A$ anomaly.

However, by coupling to the surface of the free fermion SPT we can fix this problem.  The idea is to couple the rotors $\phi$ to appropriate $U(1)_V$-breaking mass terms at the surface of the free fermion SPT, which have the property that a vortex configuration of these mass terms has $U(1)_A$ charge precisely minus the self-linking number of $d\lr{d\phi}$. A simple free-fermion calculation in QFT demonstrates that the $U(1)_V$ charge 1 mass has precisely this property, and this is a consequence of the anomaly (see Appendix \ref{app:free_fermion_vortices}).

Thus, when we couple the free fermion SPT boundary to the decorated rotor model in this way, we obtain a $U(1)_V$ breaking state with net zero $U(1)_A$ eigenvalue for each configuration of the $\phi$.  There is then no $U(1)_A$ symmetry obstruction to allowing the rotors to fluctuate. In other words, we now have an anomaly-free $U(1)_V$ spontaneous symmetry breaking phase which can then be driven into a trivial one by flucations\footnote{This argument is quite analogous to the one given in \cite{Wen_2013}, and requires more careful analysis to be conclusive.}. The fluctuation terms also have to change the fermionic state on the rotor side, by re-arranging the Kitaev chain configurations to ensure that they always conform to the vortex configurations.  We will not explicitly write down a Hamiltonian that does this, though we note that similar SPT models of Majorana decorated domain walls were constructed in $2+1$d in Ref. \cite{Tarantino}, and in $3+1$d in Refs. \cite{Wang_2018, Wang_2020}.  Because the overall $U(1)_A$ is conserved when the $\phi$ fields (and hence vortices) fluctuate, fermion parity is also conserved.  However, there are certainly fluctuations that hop odd fermion parity from the decorated rotor model to the free fermion SPT: these occur when the self-linking number of the vortex configuration (and hence its Kasteleyn-orientedness) changes.

The resulting trivial gapped ground state may be viewed (perturbatively in the strength of the $\phi$ fluctuations relative to the gap) as a superposition over all the $\phi$ field configurations of tensor product states between the lattice rotor model and the free fermion SPT.  This is our desired $U(1)_V$ and $U(1)_A$ respecting interface.  If we break all symmetries we can adiabatically continue this interface to a trivial product state, showing that the interface is topologically trivial, as desired.

\section{Discussion} \label{sec:discussion}

We have introduced the tool of symmetry disentanglers for the construction of Hamiltonian lattice models of chiral gauge theories. We have demonstrated that in 1+1D we can construct exactly-solvable Hamiltonians for a broad class of theories of $U(1)$ gauge fields coupled to Dirac fermions, subject only to the constraint of 't Hooft anomaly vanishing, thus evading fermion doubling. We have shown how to construct $U(1)_V \times U(1)_A$ symmetries in 3+1D with anomaly corresponding to four left-handed Weyl fermions with the charges eq.~\ref{eqndiraccharges}, and how to disentangle anomaly-free combinations of these symmetries.

These results in principle allow us to place the $U(1)$ hypercharge sector of the Standard Model with sterile neutrinos on the lattice. Although we are lacking an exactly-solvable $U(1)_V \times U(1)_A$ symmetric Hamiltonian giving rise to the four Weyls in the IR, we have described how to reduce this problem to constructing a trivial gapped interface between two particular free-fermion and commuting-projector Hamiltonians for the same 4+1D symmetry protected topological phase. We have given a sketch of a construction of such an interface, but filling in the details and verifying it must be left to future work.

As a byproduct of our work, the symmetry disentanglers we construct can be leveraged to produce new commuting-projector Hamiltonians for $U(1)_V \times U(1)_A$ SPT phases in 2+1D and 4+1D. The existence of these commuting-projector Hamiltonians is actually sharply constrained by various no-go theorems in models with finite-dimensional site Hilbert spaces \cite{kapustin2020local, C_Zhang_2022}. In particular, one can show that the SPT we construct in 2+1D has a non-vanishing Hall conductance, and thus cannot have a local commuting-projector model with finite-dimensional site Hilbert spaces. This is likely also true of the 4+1D SPT, demonstrating the necessity of using infinite-dimensional\footnote{This countable infinite dimensionality is distinct from the uncountable infinite dimensionality that would result from collapsing a non-compact extra dimension into $D$ dimensions. The result of the latter construction would not even be a Hilbert space.} rotor Hilbert spaces as our local degrees of freedom. Commuting-projector models for $U(1)$ SPTs in 2+1D avoiding these no-go theorems using rotors were first constructed in \cite{demarco2021commutingprojectormodelnonzero}.

A related setting to rotors is that of (Hamiltonian) Villain models, which as well as having (countably) infinite-dimensional local Hilbert spaces also have a Gauss law. Villain models are convenient for encoding the winding symmetry of the $1+1$ dimensional compact boson \cite{Cheng_2023}, as well as magnetic symmetries of abelian gauge theories. We have shown that this setting is actually equivalent to the local rotor setting. In particular, we constructed a disentangler for the Villain Gauss law, mapping the gauge-invariant Hilbert space to a local tensor product space, which may be of independent interest.

We note an important technical subtlety which holds when studying infinite-dimensional Hilbert spaces like $L^2(S^1,\mathbb{C})$. Unlike finite-dimensional systems, Hamiltonians acting on rotor Hilbert spaces are generally unbounded operators, even for finite systems. As a result, the domain of the Hamiltonian must be specified with care. This is standard for unbounded operators \cite{reed1981functional} although we usually ignore it as the domain for familiar single particle Hamiltonians is more-or-less obvious. In particular, the disentanglers we construct do not preserve the naive domain of finite angular-momentum states: finite-energy wavefunctions necessarily acquire discontinuities in the rotor variables, corresponding to twisted boundary conditions correlated across sites. While time evolution remains well defined, these features may have implications for numerical implementations and deserve further study.

\bibliographystyle{alpha}
\bibliography{main}

\appendix

\section{Chiral anomaly}\label{appchiralanom}

Let $A_V$ be a $U(1)$ gauge field and $A_A$ be a Spin$^c$ structure. The anomaly corresponding to the four left-handed Weyl fermions with the charge assignments \eqref{eqndiraccharges} may be expressed as a 4+1d Chern-Simons action
\[\label{eqnappanom}\int_{X^5} A_A (dA_V/2\pi)^2.\]
$A_A$ is a Spin$^c$ structure since a $\pi$ rotation in $U(1)_A$ is the fermion parity.

This can be confirmed by the Atiyah-Singer index theorem. Indeed, the corresponding index on a closed Spin$^c$ 6-manifold $Z^6$ is \cite{nakahara2018geometry}
\[\text{6d index}=\int_{Z^6} \frac16 {\rm Tr\ }(F/2\pi)^3 - \frac{1}{24} p_1 {\rm Tr\ }F/2\pi\]
where $F$ is the curvature of the connection on the rank 4 complex vector bundle determined by \eqref{eqndiraccharges}, i.e.
\[F = \begin{pmatrix}
    dA_V+dA_A & 0 & 0 & 0 \\
    0 & -dA_V + dA_A & 0 & 0 \\
    0 & 0 & -dA_A & 0 \\
    0 & 0 & 0 & - dA_A
\end{pmatrix}\]
This has a vanishing trace so the gauge-gravity term in the index formula does not contribute. Expanding the first term gives
\[\text{6d index}= \frac{1}{(2\pi)^3}\int_{Z^6} dA_A (dA_V)^2.\]
The free-fermion anomaly is the Chern-Simons term whose curvature gives $2\pi$ times the index, so we find the anomaly \eqref{eqnappanom} above.

\section{Commuting-projector Hamiltonians and symmetry disentanglers}\label{appcphfromdisentangler}

Let $G$ be a group and $U(g)$ a collection of (possibly not-on-site) unitary transformations on a tensor product Hilbert space\footnote{For this section we could work with either bosonic or fermionic (i.e. $\mathbb{Z}_2$-graded) Hilbert spaces. In the fermionic case, all tensor products should be $\mathbb{Z}_2$-graded tensor products. All unitaries and Hamiltonians we consider are Fermi-even.} $\cH$ (in $d=D-1$ space dimensions $\mathbb{Z}^d$) satisfying the group laws
\[U(g) U(h) = U(gh) \\ U(e) = 1\]
for all $g,h \in G$, where $e \in G$ is the identity. 

Suppose that $V(g)$ is another such collection and $W$ is a unitary transformation on the doubled Hilbert space $\cH \otimes \cH$ such that for all $g \in G$
\[\label{eqninvertibleanom}W(U(g) \otimes V(g))W^\dagger = U_0(g) \otimes U_0(g).\]
where $U_0(g)$ are on-site $G$ symmetries acting on $\cH$. We will want to assume some kind of locality-preserving property for $W$, which will control the range of the Hamiltonians we will construct. For the symmetry disentanglers we construct, $W$ is a constant-depth circuit, which will lead to Hamiltonians whose terms have a bounded size. We will assume this below.

In the text, for $G = U(1)$ and $U(\theta)$, $\theta \in U(1)$ given by a combination of axial and vector rotations (for bosons or fermions) in 1+1D or 3+1D with charges $q_V^\alpha$, $q_A^\alpha$, we have constructed a disentangler $W$ which applies to $U(\theta) \otimes V(\theta)$, where $V(\theta)$ is any combination of axial and vector rotations with the opposite anomaly to $U(\theta)$. This is a special case of the disentanglers we have constructed for anomaly free subgroups, since $U\otimes V$ certainly has all cancelling anomalies.

\textit{Definition:} Given $\{U(g),V(g)\}_{g \in G}$ and $W$, we construct a corresponding Hamiltonian $H_{SPT}$ in $d{+}1$ spatial dimensions with an on-site symmetry as follows.

We define $H_{SPT}$ on a lattice $\mathbb{Z}^{d{+}1}$ which in the $d{+}1$st direction looks like stacked layers of our $d$-dimensional Hilbert space $\cH$, which we write as $\cH_j$, with $j \in \mathbb{Z}$ labeling the $d{+}1$st coordinate. Let $W_{j,j+1}$ be the constant-depth circuit in \eqref{eqninvertibleanom} above acting on $\cH_j \otimes \cH_{j+1}$. There is likewise a $\tilde W_{j,j+1}$ acting on $\cH_{j} \otimes \cH_{j+1}$ such that
\[\tilde W(V(g) \otimes U(g))\tilde W^\dagger = U_0(g) \otimes U_0(g).\]
Consider the ``swindle''
\[S = \prod_k \tilde W_{2k+1,2k+2} \prod_j W_{2j,2j+1}^\dagger.\]
It can be checked that $S$ commutes with the on-site symmetry $\bigotimes_{j \in \mathbb{Z}} U_0(g)$. We may suppose (adding ancillas if we must) that $U_0(g)$ admits a symmetric Hamiltonian $H_0$ having a unique product state ground state and with each term a single site projector. We thus define
\[H_{SPT} := S H_0 S^\dagger.\]
This is a commuting-projector Hamiltonian with bounded terms, since $S$ sends the single-site commuting-projector terms of $H_0$ to commuting projectors of bounded size. Futhermore, since $S$ commutes with $\bigotimes_{j \in \mathbb{Z}} U_0(g)$, as does $H_0$, $H_{SPT}$ has the on-site symmetry $\bigotimes_{j \in \mathbb{Z}} U_0(g)$.

\textit{Claim:} Symmetric boundaries of $H_{SPT}$ are in a one-to-one correspondence with $d$-dimensional Hamiltonians with the not-on-site symmetry $U(g)$, up to the addition of ancillas carrying on-site $G$ symmetry.

\begin{proof} Let $H_{R}$ be a Hamiltonian defined on $\cH_R=\bigotimes_{j \ge 0} \cH_j$ which restricts to $H_{SPT}$ a distance $l$ into the bulk and which enjoys the symmetry
\[\label{eqnboundarysymmetry}U_{0,R}(g)=\bigotimes_{j \ge 0} U_0(g).\]
We consider
\[S_{R} = \prod_{k \ge 0} \tilde W_{2k+1,2k+2}\prod_{j \ge 0} W^\dagger_{2j,2j+1}.\]
Applied to the above,
\[\label{eqndisentangledsptwithboundary}H_{R,\text{disentangled}} := S_{R}^\dagger H_{R} S_{R} = H_\partial + \sum_{j\geq l+3} H_0^j,\]
where $H_0^{j}$ is the sum of the decoupled fully gapping projectors $H_0$ over the $j$th layer.  $H_{R,\text{disentangled}}$ is thus  a Hamiltonian on $\cH_R=\bigotimes_{j \ge 0} \cH_j$ which is equal to some boundary term $H_\partial$, supported on $\cH_\partial = \bigotimes_{j=0}^{l+2} \cH_j$, plus decoupled onsite projectors $H_0$ for $j \geq l+3$. $H_{R,\text{disentangled}}$ further has
\[\label{eqndisentangledboundarysymmetry}S_{R}^\dagger U_{0,R}(g) S_{R} = U(g) \otimes \left( \bigotimes_{j \ge 1} U_0(g)\right)\]
as a symmetry. In particular, $H_\partial$ has
\[\label{eqn_simplified_boundary_symmetry}U(g) \otimes \bigotimes_{j=1}^{l+2} U_0(g)\]
as a symmetry. This concludes the construction of the $d$-dimensional Hamiltonian with not-on-site symmetry.

Conversely, suppose we are given $H_\partial$ on $\cH_\partial = \bigotimes_{j=0}^{l+2} \cH_j$ with the symmetry in eq. \ref{eqn_simplified_boundary_symmetry} above. We can then reconstruct $H_{R,\text{disentangled}}$ and $H_R$ via \eqref{eqndisentangledsptwithboundary}:
\[H_R = S_R\left(H_\partial + \sum_{j\geq l+3} H_0^j\right)S_R^\dagger\]
This inverts the construction above.
\end{proof}

\textit{Claim:} If the original not-on-site symmetry $U(g)$ is disentanglable, we may construct a trivial symmetric boundary of $H_{SPT}$.

\begin{proof}So we suppose there is a constant-depth circuit $C$ such that for all $g \in G$
\[C^\dagger U(g) C = U_0(g).\]
For example, this is the case when the $U(g)$ is a combination of our rotor model axial and vector rotations with vanishing anomaly, in which case circuits that made the symmetry on-site were constructed in the main text.  With this assumption, we will construct a trivial symmetric boundary for the SPT Hamiltonian. Let us take $C$ to act only on the first tensor factor $\cH_0$ of $\cH_R=\bigotimes_{j \ge 0} \cH_j$ (i.e. $j=0$). We consider
\begin{align}
H_{R,\text{trivially gapped}} = S_R C \left(\sum_{j=0}^{\infty} H_0^{j} \right)C^\dagger S_R^\dagger,
\end{align}
where $H_0^j$ is the sum of the onsite $H_0$ terms over all the sites of layer $j$.  This is a commuting-projector Hamiltonian restricting to $H_{SPT}$ on $\bigotimes_{j \ge 3} \cH_j$ and having the onsite symmetry
\[C^\dagger S_{R}^\dagger U_{0,R}(g) S_{R} C = U_{0,R}(g).\]
(cf. \eqref{eqndisentangledboundarysymmetry}). It furthermore has a unique gapped ground state, since it is the unitary conjugate of a sum of trivial projectors that gap out everything, and so we regard it as a trivial symmetric boundary of $H_{SPT}$.

\end{proof}

Note also that the above claim applies to the situation where a subgroup $G' \subset G$ of the original not-on-site symmetry is disentanglable.  In that case, the argument leads to a $G'$-symmetric trivial gapped boundary for the $G'\subset G$ SPT.  Also, as mentioned above, this entire discussion extends to the commuting projector fermionic SPT phases used in this paper.

\section{Cochain formalism}\label{appcochains}

In this appendix we review some basic facts about simplicial (co)chains, (co)cycles, and (co)homology. A basic reference for this material is \cite{hatcher2005algebraic} or \cite{thorngren2018combinatorial} for some physics context.

A geometric $k$-simplex is a space homeomorphic to the convex hull of $k+1$ non-hyperplanar points in $\mathbb{R}^{k+1}$, such as the unit coordinate vectors $\hat e_1,\ldots, \hat e_{k+1}$. For $k = 0$ it is a point, $k = 1$ is an interval, $k = 2$ is a triangle, $k = 3$ is a tetrahedron and so on. Note that the $j$-faces of such a geometric $k$-simplex, meaning the convex hull of $j+1$ of the vertices $\hat e_1,\ldots, \hat e_{k+1}$, is also a geometric $j$-simplex.

It is convenient to think about topological spaces as unions of geometric simplices joined along their $j$-faces. We call such a space a simplicial complex.

Let $X$ be a simplicial complex. We define $C_k(X)$ to be the free abelian group generated by the oriented $k$-simplices $\sigma^k$ of $X$, modulo the relation
\[ [\bar\sigma^k] = -[\sigma^k],\]
where $[\sigma^k]$ denotes the generator corresponding to the $k$-simplex $\sigma^k \subset X$, and $\bar \sigma^k$ is $\sigma^k$ with the opposite orientation. Concretely, elements of $C_k(X)$ are finite sums
\[\sum_{i=1}^n a_i [\sigma_i^k],\]
where $a_i \in \mathbb{Z}$. These are called (integer) $k$-chains.

We define a homomorphism $\partial:C_k(X) \to C_{k-1}(X)$ on generators as
\[\partial [\sigma^k]=\sum_{\tau \in \partial \sigma} [\tau^{k-1}],\]
where the sum is over the boundary $k-1$-simplices of $\sigma^k$, with their orientation inherited from that of $\sigma^k$. This satisfies $\partial^2 = 0$.

Although it is not so important for us here, one then defines the subgroup of $k$-cycles $Z_k(X)={\rm Ker}(\partial) \subset C_k(X)$ as well as the subgroup of $k$-boundaries $B_k(X)={\rm Im}(\partial) \subset C_k(X)$, then $B_k(X) \subset Z_k(X)$ and so we may define the $k$th homology
\[H_k(X) = \frac{Z_k(X)}{B_k(X)}.\]

Let $A$ be an abelian group. We define $C^k(X,A)$ to be the abelian group of homomorphisms from $C_k(X)$ to $A$. For $\alpha \in C^k(X,A)$, $Y \in C_k(X)$, we adopt the notation
\[\int_Y \alpha := \alpha(Y).\]
Elements $\alpha \in C^k(X,A)$ are called $A$-valued $k$-cochains. We can express a general such cochain as a finite sum
\[\sum_{i=1}^n a_i \delta_{\sigma_i^k},\]
where $a_i \in A$ and $\delta_{\sigma_i^k}$ is the $\mathbb{Z}$-valued $k$-cochain which is 1 on the oriented $k$-simplex $\sigma_i^k$ and 0 on all other $k$-simplices.

$\partial$ induces a map $d:C^k(X,A) \to C^{k+1}(X,A)$ meaning for $Y \in C_{k+1}(X)$, $\alpha \in C^k(X,A)$,
\[\int_Y d\alpha := \int_{\partial Y} \alpha.\]
Note that this implies $\int_Y d\alpha = 0$ when $\partial Y = 0$. This map satisfies $d^2 = 0$.

We may then define the group of $k$-cocycles $Z^k(X,A) = {\rm Ker}(d) \subset C^k(X,A)$, the group of exact $k$-cocycles $B^k(X,A) = {\rm Im}(\partial) \subset C^k(X,A)$, and the $k$th cohomology
\[H^k(X,A) = \frac{Z^k(X,A)}{B^k(X,A)}.\]

Now suppose $A = R$ is a commutative ring and $X$ has a branching structure, which is an ordering of all the 0-simplices of $X$. We define the cup product $\cup:C^k(X,R) \otimes_R C^j(X,R) \to C^{k+j}(X,R)$ given $\alpha \in C^k(X,R)$, $\beta \in C^j(X,R)$, on $\sigma^{k+j+1}$ as follows. Given the branching structure, we can order the $k+j+1$ vertices of $\sigma^{k+j+1}$ as $v_0 < v_1 < \cdots < v_{k+j}$. Let $\tau^k$ be the $k$-simplex of $\sigma^{k+j+1}$ containing the vertices $v_0,\ldots, v_k$ and let $\rho^j$ be the $j$-simplex of $\sigma^{k+j+1}$ containing the vertices $v_k,\ldots, v_{k+j}$. We define
\[\int_{\sigma^{k+j+1}} \alpha \cup \beta := \alpha(\tau^k) \beta(\rho^j).\]
This satisfies the Leibniz rule
\[d(\alpha \cup \beta) = (d\alpha) \cup \beta + (-1)^k \alpha \cup (d\beta).\]
This implies that the cup product is well-defined on cohomology, making $\bigoplus_k H^k(X,R)$ a graded ring. It is in fact graded-commutative, but the product on $\bigoplus_k Z^k(X,R)$ is only graded-commutative up to correction terms. See \cite{thorngren2018combinatorial} for more details.

\section{Disentangling Villain Gauge Theories}\label{appdisentanglingvillaingaugetheory}

For rotors, we described in \eqref{eqnvillaindisentangler} a Villain disentangler mapping the Hilbert space of states satisfying the Villain condition to a tensor product Hilbert space. The same disentangling transformation can be applied to Villain $U(1)$ gauge theories. This may be of independent interest.

Villain $U(1)$ gauge theory can be thought of as resulting from gauging the $\mathbb{Z}$ 1-form symmetry of $\mathbb{R}$ gauge theory. We consider the Villain Hilbert space as embedded in a ``big'' vector space consisting of wavefunctions $\Psi(\{a_e\}_e,\{c_p\}_p)$ where $a_e \in \mathbb{R}$ are associated to edges $e$ of the lattice and $c_p \in \mathbb{Z}$ are associated to plaquettes $p$. We can also write $a \in C^1(X,\mathbb{R})$, $c \in C^2(X,\mathbb{Z})$ in the cochain formalism.

The Hilbert subspace of physical states of the big vector space are those invariant under two gauge transformations. One is the $\mathbb{R}$ 0-form gauge transformation:
\[a \mapsto a + dg\]
where $g \in C^0(X,\mathbb{R})$. The other is the 1-form Villain transformation
\[a \mapsto a + n \\
c \mapsto c + dn\]
where $n \in C^1(X,\mathbb{Z})$.

Let $\chi_p$ be the conjugate variable to $c_p$, satisfying
\[e^{i \chi_p} |c_p\rangle = |c_p + 1\rangle.\]
We define
\[C_e = \exp\left( i \lr{a_e} \sum_{p, e \subset \partial p} \pm \chi_p\right)\]
where the sum is over plaquettes $p$ with $e \subset \partial p$ sign is either $+$ if the orientation of $p$ agrees with $e \in \partial p$ or $-$ otherwise. These are the same signs produced when computing $d\delta_e$, where $\delta_e \in C^1(X,\mathbb{Z})$ is the 1-cochain which is $1$ on the edge $e$ and zero on all others.

The $C_e$ are invertible and commute for separate edges. We can thus consider the constant-depth circuit $C = \prod_e C_e$. We have
\[\exp\left( i \lr{a_e + n_e} \sum_{p, e \subset \partial p} \pm \chi_p\right) \\ = \exp\left( i \lr{a_e} \sum_{p, e \subset \partial p} \pm \chi_p\right) \exp\left( i n_e \sum_{p, e \subset \partial p} \pm \chi_p\right)\]
so $C$ transforms the 1-form gauge transformations into
\[a \mapsto a + n\\
c \mapsto c.\]
In the image of the physical Hilbert space we can thus regard $a_e$ as a 1-periodic variable so $a \in C^1(X,U(1))$. We can choose a branch of the log and express $a_e \in [0,1)$.

For 0-form transformations the rule is more complicated. We will have
\[\exp\left( i \lr{a_e + (dg)_e} \sum_{p, e \subset \partial p} \pm \chi_p\right) \\ =  \exp\left( i \lr{a_e} \sum_{p, e \subset \partial p} \pm \chi_p\right)  \exp\left( i (\lr{a_e + (dg_e)}-\lr{a_e}) \sum_{p, e \subset \partial p} \pm \chi_p\right).\]
Thus
\[a \mapsto a + dg \\
c \mapsto c - ds(a,g)\]
where
\[s(a,g)_e = \lr{a_e + (dg)_e} - \lr{a_e}\]
defines an integer 1-cochain.

\section{Vortices in the $U(1)_V$ breaking phase - free fermion computation} \label{app:free_fermion_vortices}

In this appendix we discuss the coupling of free fermions to the symmetry broken phase of the rotors, which is meant to balance the $U(1)_A$ charge of the vortices. The behavior of these $U(1)_A$ charges actually depends only on the anomaly \eqref{eqnappanom}, which we will show below.

Our free fermion SPT surface has $4$ Weyl fermions, with $U(1)_V$ and $U(1)_A$ charges given by eq.~\ref{eqndiraccharges}.  For the following discussion, we will find it useful to pair up the first and third Weyl fermion into a Dirac fermion $\psi'$ and the second and fourth Weyl fermion into a Dirac fermion $\psi''$.  We do this by transforming the handedness of the third and fourth Weyls and flipping all the $U(1)$ quantum numbers.  Note that this is different from the pairing discussed in the context of pumping in the main text, but we are free to pair up our Weyls as we like.  With this pairing, $U(1)_A$ acts as a diagonal vector-like symmetry $U(1)_V' \times U(1)_V''$, and $U(1)_V$ acts as chiral rotation in $\psi'$, i.e. only on the left handed part of $\psi'$, and the opposite chiral rotation in $\psi''$.  

The idea is to couple the charge 1 order parameter of the $U(1)_V$ symmetry breaking phase of the rotors to the Dirac masses for $\psi'$ and $\psi''$ such that the microscopic $U(1)_V$ (and $U(1)_A$) symmetry acting on both the decorated rotors and the surface fermions is preserved. That is, on the lattice we express the Dirac masses as local operators $\mathcal{O}_{1,j}$, $\mathcal{O}_{2,j}$ supported near site $j$, forming a $U(1)_V$ charge $-1$ doublet, and write the $U(1)_V \times U(1)_A$ symmetric coupling
\[H_{coupling} = - M \sum_j \mathcal{O}_{1,j}\cos \phi_j + \mathcal{O}_{2,j} \sin \phi_j.\]
With a fixed $\phi$ configuration, this effectively determines the mass of the Dirac fermions as
\begin{align}
m' + i m'_5 &= Me^{i\phi} \\
m'' + i m''_5 &= Me^{-i\phi}
\end{align}
where we collect the masses $m \bar \psi \psi$ and $im_5 \bar \psi \gamma^5 \psi$ into a complex mass. This will gap out the fermions leaving only the $U(1)_V$ Goldstone modes at low energies. This will then be a mundane $U(1)_V$ spontaneous symmetry breaking phase with no obstruction to being driven into a trivial phase by fluctuations. 

As described in the main text, without this coupling the vortices in $\phi$ carry $U(1)_A$ charge depending on their self-linking, and so fluctuations without this coupling will generically change the $U(1)_A$ charge. Let us describe how with this coupling above that the change in $U(1)_A$ charge is compensated by that of the fermions $\psi'$, $\psi''$.

It is simplest to first analyze the situation of a single infinite rotationally symmetric vortex along the $z$-axis.  Let $r = \sqrt{x^2+y^2}$ and $\theta = \tan^{-1} (y/x)$ be polar coordinates.  This problem was solved for a single Dirac fermion by \cite{JackiwRossi, Weinberg_vortex, CallanHarvey_vortex}.  The solution is a chiral zero mode bound to the core of the vortex.  The wavefunction for the $k_z=0$ consists of a decaying cylindrically symmetric profile away from the $z$-axis, a factor of $e^{i\theta}$ times a certain fixed spinor.  In our situation, we have two Dirac fermions $\psi'$ and $\psi''$, and by virtue of our choice of mass term, their vortices have opposite vorticity, which means that the solutions are counter-propagating.  To fully gap the vortex we now need to introduce a term that couples $\psi'$ and $\psi''$ and gaps out the vortex.  It is here that we need to break rotational symmetry around $z$, effectively by choosing a framing of $z$.  Actually, assuming without loss of generality that our vortex lines are present only on the dual lattice of the rotor model simplicial decomposition, we can obtain such a framing canonically from the choice of branching structure.  This framing comes from the Morse flow, in the geometric realization of simplicial cup products \cite{Thorngren2018_thesis, tata2020geometricallyinterpretinghighercup}.  The self linking number of the vortex configuration, as computed by the chochain cup product formula on the simplicial side, is then equal to the linking number between the vortex configuration and its pushoff along the Morse flow vector.

We will see below that a unit winding of the framing will bind unit $U(1)_A$ charge. By virtue of the $U(1)_A$ charge being a topological invariant we can always unwind a complicated vortex configuration to one where it looks like a long straight vortex, which will lead to the $U(1)_A$ charge being just the self-linking of this vortex, completing the argument.

This property of $U(1)_A$ charge determined by the self-linking number actually follows directly from the anomaly. Suppose we have a 3+1D system with symmetry $U(1)_V \times U(1)_A$ (with fermion parity a $\pi$ rotation in $U(1)_A$) and the anomaly \eqref{eqnappanom} corresponding to the four Weyl fermions \eqref{eqndiraccharges}
\[\int_{X^5} A_A (dA_V/2\pi)^2.\]
We consider breaking the $U(1)_V$ symmetry by a charge-1 order parameter. We apply the symmetry-breaking long-exact-sequence (SBLES) of \cite{debray2024longexactsequencesymmetry} to analyze the vortices and show that they bind a $U(1)_A$ charge equal to their self-linking number.

Following \cite{debray2024longexactsequencesymmetry}, if we consider a perfectly straight vortex along the $z$ axis, with symmetry breaking field rotationally symmetric around this axis, then although $U(1)_V$ is broken, there is an unbroken symmetry $U(1)_{\tilde V}$ combining $U(1)_V$ with spatial rotations in the $xy$ plane. Note that a $\pi$ $U(1)_{\tilde V}$ rotation amounts to the fermion parity, so the symmetry group of the straight vortex is
\[(U(1)_A \times U(1)_{\tilde V})/\Z_2.\]
By \cite{debray2024longexactsequencesymmetry}, this symmetry acts anomalously on localized modes along the vortex, and the anomaly may be computed from \eqref{eqnappanom} to be
\[\label{eqnreducedanomstraightvortex}\frac{1}{4\pi} A_A dA_{\tilde V}\]
(The factor of $1/2$ comes because we have re-normalized $A_{\tilde V}$ relative to $A_V$ so that $A_{\tilde V}$ is a Spin$^c$ structure.) We can change to the usual basis
\[A_A = A_L - A_R \\
A_{\tilde V} = A_L + A_R,\]
so anomaly becomes
\[\frac{1}{4\pi} A_L dA_L - \frac{1}{4\pi} A_R dA_R.\]
We recognize this as the anomaly corresponding to the vector and axial symmetries of a single Dirac fermion, consistent with the free-fermion analysis.

Now we want to consider deformations from the straight vortex. These deformations will break the rotational symmetry and hence $U(1)_{\tilde V}$, which generically will be broken all the way down to fermion parity. Following \cite{debray2024longexactsequencesymmetry} we introduce a $2\pi$-periodic spurion field $\theta$ which has the minimal charge 2 under $U(1)_{\tilde V}$ to capture this symmetry breaking. The anomaly \eqref{eqnreducedanomstraightvortex} becomes a topological term
\[\frac{1}{2\pi} A_A d\theta.\]
This indicates that a winding number $w$ of $\theta$ contributes an axial charge of $w$ for the vortex states.

Since the spurion $\theta$ transforms as a normal vector under the broken rotational group, we can interpret it as a framing for the vortex. The anomaly thus shows that when the framing changes by $k$ units, the axial charge must also change by $k$ units.

\end{document}